\documentclass[10pt]{article}

\usepackage[a4paper,top=23mm,bottom=23mm,left=32mm,right=32mm]{geometry}

\usepackage{amsmath,amssymb}
\usepackage{array}
\usepackage{graphicx}
\usepackage{float}
\usepackage[numbers,sort&compress,round]{natbib}
\usepackage{color}
\usepackage{placeins}
\usepackage{caption}
\usepackage[
    unicode,
    colorlinks=true,
    urlcolor=blue,
    linkcolor=blue,
    citecolor=blue
]{hyperref}
\usepackage{lmodern}
\usepackage[most]{tcolorbox}
\usepackage{tikz}
\usepackage{varwidth}
\usepackage{wrapfig}
\usepackage{capt-of}
\usepackage{paracol}
\usepackage{soul}
\usepackage{units}
\usepackage{xcolor}
\definecolor{myOrange}{HTML}{F4A81E}
\definecolor{myBlue}{HTML}{4787C7}
\definecolor{MyOrange}{HTML}{FCBF43}
\definecolor{MyGreen}{HTML}{67C189}

\newcommand{\rev}{\textcolor{black}}

\setcitestyle{round}

\newcommand{\um}{$\mu$m }

\newcommand{\Leq}{$L_{\rm eq\,}$}

\begin{document}

\begin{center}

{\Large\bf %
Quantum trapping and rotational self-alignment in triangular Casimir microcavities
}
\vspace{3ex}

Betül Küçüköz,$^{1,\ast}$ Oleg V. Kotov,$^{1,\ast}$ Adriana Canales,$^{1,\ast}$ Alexander Yu. Polyakov,$^{1}$ Abhay V. Agrawal,$^{1}$ Tomasz J.\ Antosiewicz,$^{2,1}$ and Timur O. Shegai$^{1,\dagger}$

\vspace{1ex}

\emph{
$^{1}$Department of Physics, Chalmers University of Technology, 412 96, Gothenburg, Sweden \\
$^{2}$Faculty of Physics, University of Warsaw, Pasteura 5, 02-093 Warsaw, Poland \\
$^{\ast}$These authors contributed equally to this work. \\
$^{\ddagger}$email: \href{mailto:timurs@chalmers.se}{timurs@chalmers.se} \\
}

\end{center}

\rev{Casimir torque -- a rotational motion driven by zero-point energy minimization -- is a problem that attracts notable research interest. Recently, it has been realized using liquid crystal phases and natural anisotropic substrates. However, for natural materials, substantial torque occurs only at van der Waals distances of $\sim 10$ nm.
Here, we employ Casimir self-assembly with triangular gold nanostructures for rotational self-alignment at truly Casimir distances (100 -- 200~nm separation). The interplay of repulsive electrostatic and attractive Casimir potentials forms a stable quantum trap, giving rise to a tunable Fabry-Pérot microcavity. This cavity self-aligns both laterally and rotationally to maximize area overlap between templated and floating flakes. The rotational self-alignment is sensitive to the equilibrium distance between the two triangles as well as their area, offering possibilities for active control via electrostatic screening manipulation. Our self-assembled Casimir microcavities present a versatile and tunable platform for nanophotonic, polaritonic, and optomechanical applications.}





\vspace{1cm}
\noindent
{\Large{\textbf{Introduction}}}
\vspace{0.5cm}

\noindent
The quantum nature of van der Waals forces was initially revealed by London \cite{london1937} and subsequently generalized by Casimir and Polder \cite{CasimirPolder1948}. Furthermore, Casimir extended this concept to describe the attraction between two ideal mirrors \cite{casimir1948}, which brought a quantum electrodynamics effect to the macroscopic scale. Following this, Lifshitz and co-authors developed a theory that allowed for the calculation of the Casimir effect between arbitrary planar mirrors, relying on the classical optical response of materials~\cite{lifshitz1956,Lifshitz-Dzyaloshinskii-Pitaevski1961}. This advancement paved the way for predicting both repulsive~\cite{,Munday2009,Munday2010} and lateral Casimir forces~\cite{chen2002demonstration,chen2002experimental,rodrigues2006lateral} in various contexts.

The introduction of asymmetry allows extending the Casimir effect to a new degree of freedom -- rotation. In this respect, Casimir torque led by quantum fluctuations was first predicted by Kats \cite{kats1971} and Parsegian and Weiss \cite{Parsegian1972Torque} when considering dielectric media with in-plane anisotropy at short separation distances. Later, the approach was generalized to longer distances by Barash \cite{Barash1978}. However, the predicted torque was small and turned out to be challenging to observe. The first Casimir torque measurements were reported only recently~\cite{Munday2018Torque,Rev2022Torque}. In these measurements, the air gap was replaced with a solid isotropic interlayer, which helped support two anisotropic materials at a fixed distance and in a parallel configuration. The torque was measured by optical characterization of the twist of a liquid crystal, which acted as one of the birefringent bodies. Such a setup allowed controlling the sign and strength of the torque by choosing an anisotropic substrate material and varying the interlayer thickness, but did not allow to observe the torque directly, since the rotation was hidden inside the liquid crystal. Furthermore, the torque between two media with artificial in-plane anisotropy, such as lamellar gratings, was predicted to be substantially greater than in natural anisotropic materials \cite{guerout2015,AntezzaTorque2020}. Particularly, Guerout \textit{et al.} \cite{guerout2015} obtained the torque per unit area for the infinite gratings and accounted for the finite-size effects using the overlap area approximation, which works well when the lateral size of the gratings is much larger than their characteristic length scales and the gap between them. Recently, Antezza \textit{et al.} \cite{AntezzaTorque2020} calculated the Casimir torque between finite-sized metallic gratings beyond the overlap approximation. It turned out that a finite number of gratings periods leads not only to oscillations of the torque direction but also to a much larger magnitude of the torque than intuitively expected. Moreover, due to a critical zero-order geometric transition between a 2D- and a 1D-periodic system, the torque per unit area can reach extremely large values, increasing without bounds with the size of the system, which paves the way to observation of the torques at truly Casimir distances $\sim$ 100~nm and beyond.
 
Importantly, an alternative way to observe Casimir torques is offered by the optical levitation of anisotropic dielectric nanoparticles (silica in particular) in vacuum \cite{xu2017,ahn2018,ahn2020}. This method has ultra-high sensitivity but requires high vacuum conditions and works only with relatively small nanoparticles that can be captured using conventional optical tweezers. Therefore, this method does not allow direct imaging of the Casimir torque in an optical microscope but instead relies on polarization-dependent readout. 

We note that the aforementioned works primarily focus on Casimir interactions in dry or vacuum environments. However, the introduction of liquids can be beneficial -- in particular, repulsive Casimir forces have been successfully demonstrated using a combination of high refractive index liquid bromobenzene interfaced between two solids -- gold and glass substrates \cite{Munday2009}, as well as in gold-ethanol-teflon systems \cite{ZhangScience2019}. Furthermore, the liquid environment allows to use colloids as a platform for studying Casimir interactions. The theoretical framework for describing colloidal interactions in solution is often based on the DLVO theory, which typically involves electrostatic stabilization to balance attractive van der Waals forces \cite{derjaguin1941acta,verwey1947theory}. Recently, the DLVO theory has been extended to account for retardation effects \cite{biggs1994measurement,Nature2021}.

Van der Waals and Casimir interactions in colloidal solutions play an important role not only in their stability but also in self-assembly \cite{Grzybowski2002,Chiang2007,batistareview2015,ZhangScience2019, esteso2019casimir,ge2020tunable,schmidt2023tunable}. Recently, the formation of self-assembled, stable Fabry-Pérot (FP) cavities has been achieved through a combination of attractive Casimir and repulsive electrostatic interactions \cite{Nature2021}. This development opens up new possibilities for strong light-matter interactions and highlights the intrinsic relationship between the original Casimir problem and the planar microcavity problem. Moreover, the electrostatic force in the system can be actively modified, enabling the tuning of the FP resonance within a certain range without compromising cavity stability. The repulsive electrostatic force in the Casimir microcavity can be controlled by adjusting the ion concentration of the aqueous solution. This delicate balance between the attractive Casimir force and the repulsive electrostatic force not only stabilizes the FP cavity in the vertical direction but also enables the formation of laterally stable structures.

\begin{figure}
\includegraphics[width=1\textwidth]{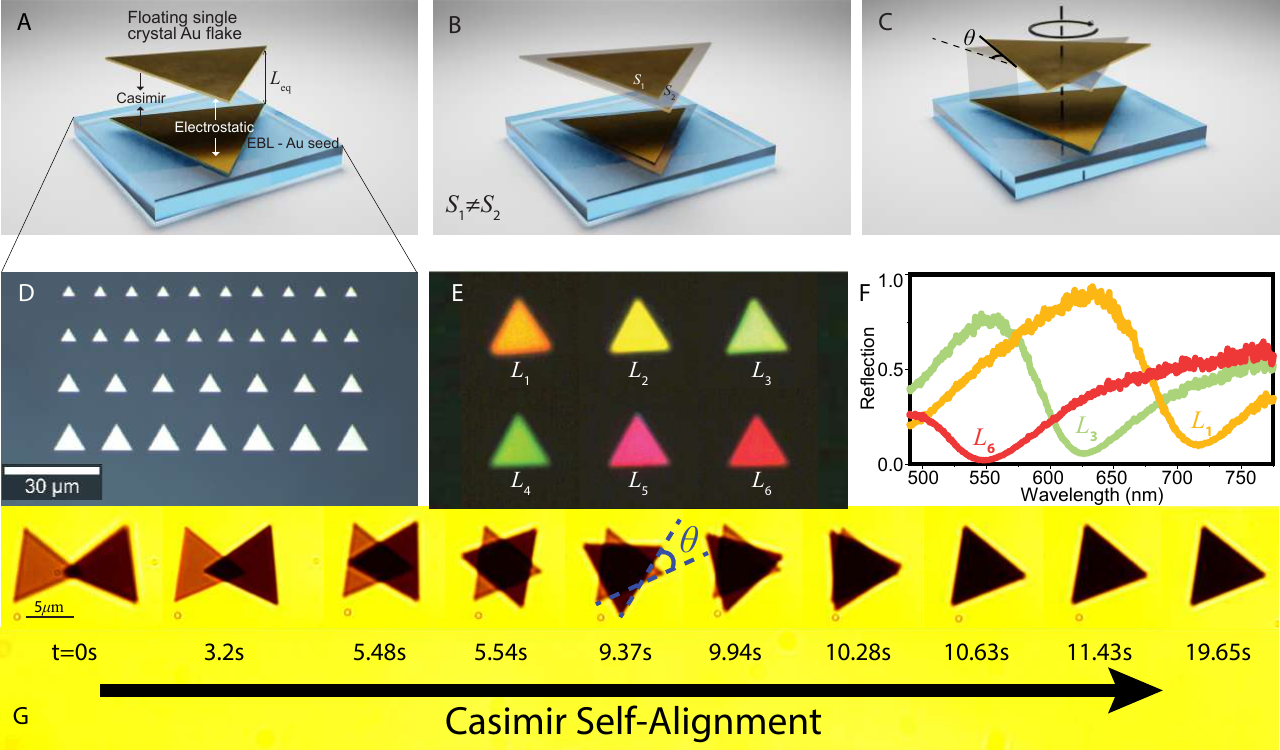}
\caption{\textbf{Self-assembled triangular Casimir microcavities composed through a combination of top-down lithography and colloidal chemistry.}  \rev{(\textbf{A})} Schematic illustration of the triangular seed approach on the glass substrate and Casimir self-assembly forming a stable Fabry-Pérot microcavity.  \rev{(\textbf{B})} The seed approach allows exploring the effect of overlap area on Casimir self-alignment and its stability to thermal fluctuations.  \rev{(\textbf{C})} Illustration of the rotation angle ($\theta$) variation as the flakes self-align. The angle $\theta$ is defined as the angle between the edges of the seed and the floating flake. When the floating flake is fully aligned to the seed $\theta = 0$.  \rev{(\textbf{D})} Optical bright field image of the templated Au seed arrays on a glass substrate with $a$ = 4, 5, 7, and 10 $\mu$m edge lengths, respectively.  \rev{(\textbf{E})} True-color reflectivity images of self-aligned Casimir microcavities using 5 \um seeds recorded at distinct $L_{\rm eq}$ values, denoted as $L_{\rm i}$ for ${\rm i=1,2,...,6}$.  \rev{(\textbf{F})} Reflection spectra of the self-aligned Fabry-Pérot microcavities used to determine $L_{\rm eq}$ by the transfer-matrix method. $L_1 > L_3 > L_6$, i.e. $L_1=199$, $L_3=161$, and $L_6=123$~nm, respectively.  \rev{(\textbf{G})} Selected video frames of the self-alignment process. The floating flake initially approaches the seed in the opposite orientation and begins rotating around $t=9.37$ s. The rotation continues until the flakes are fully aligned, after which the configuration remains stable.}
\label{fig:Fig1}
\end{figure}

The Casimir self-assembly approach offers numerous advantages, but it also presents certain challenges. These challenges include slow diffusion-limited cavity formation, irreproducibility, unscalable fabrication, and a lack of integration with microfluidics. One potential solution to address these challenges is the use of templated structures. Indeed, substrates that are templated with various nanostructured patterns are commonly employed in self-assembled colloidal systems and plasmonic arrays \cite{Xia2009templated, Mirkin2014templated, Fery2014templated, LizMarzan2021templated}. Implementing this approach could help overcome several problematic issues on the road towards realizing asymmetry-induced Casimir torques in liquid environments and scalable as well as stable formation of Casimir microcavities in vertical, horizontal, and rotational domains. However, to date, the templated self-assembly approach has not been explored for Casimir self-assembly and Casimir torques. 

Here, we introduce template-assisted systems for the self-assembly of Casimir microcavities, where patterned metallic surfaces on the substrate are crucial for achieving lateral Casimir forces and self-alignment through Casimir torque. We specifically chose an equilateral triangle geometry to realize the Casimir torque, as triangles' symmetry provides the highest possible torque among other equilateral microstructures. We investigate the self-alignment of triangular nanoflakes in an aqueous solution at room temperature by monitoring the rotational motion induced by the lateral Casimir force, which strives to maximize the overlap area between the top colloidal flake and the bottom templated flake on the substrate. This rotational effect is substantial and can be observed in real time using an optical microscope. Additionally, we study the influence of thermal fluctuations, the separation distance between the flakes, and their areas on the stability of the obtained microcavities. Our approach not only enables the scalability of the Casimir self-assembly and self-alignment but also paves the way for the integration of Casimir torque effects with colloidal science, nanophotonics, polaritonics, and self-assembly. On a broader account, it is important to mention that Casimir and Casimir torque effects studied here could be relevant for the design of micro-electromechanical devices~\cite{MEMS} and in future devices relying on an efficient and contactless transfer of angular momentum~\cite{sanders2019,golestanian2007,lombardo2008,shajesh2008}.

\vspace{1cm}
\noindent
{\Large{\textbf{Results and Discussion}}}
\vspace{0.5cm}

\noindent
The process of formation of stable Casimir microcavities in aqueous solution \cite{Nature2021} can be substantially enhanced and more accurately controlled by combining top-down nanopatterned gold areas (referred to as ``seeds") on the glass substrate and floating Au flakes in the solution (Fig. \ref{fig:Fig1}\rev{A}). This approach allows to control the density, size, and shape of the seeds, which is unavailable for the previous method. Over time, the floating Au flakes, playing the role of ``micromirrors", diffuse towards the seeds and form dimers due to lateral Casimir forces. Moreover, if the seeds are triangular, the floating flakes not only form stable cavities but also geometrically align with the seeds to maximize the overlap area (and hence minimize the Casimir potential, see Fig. \ref{fig:Fig1}\rev{C}).

In this work, we focus specifically on equilateral triangles (Fig. \ref{fig:Fig1}\rev{A}). Experiments are designed with two main components: (\textit{i}) glass substrates with the precisely fabricated seeds produced by electron beam lithography (EBL), and (\textit{ii}) single crystal Au flakes in aqueous solution produced by wet chemical synthesis \cite{anilinesynt2006}. Triangular Au seeds with lateral dimensions in the 4 -- 10 \um range and 20 nm heights are shown in Fig. \ref{fig:Fig1}\rev{D} (also see Methods). The seed approach allows us to control the Casimir force on the floating Au flake. Since it is difficult to control the colloidal growth with high precision \cite{anilinesynt2006}, chemically synthesized Au flakes are typically obtained in a range of sizes, thicknesses, and shapes. To simplify the self-alignment problem, we therefore preselect only equilateral triangle flakes of appropriate size by dragging them with optical tweezers to the seeds.  

\begin{figure}
\includegraphics[width=0.9\textwidth]{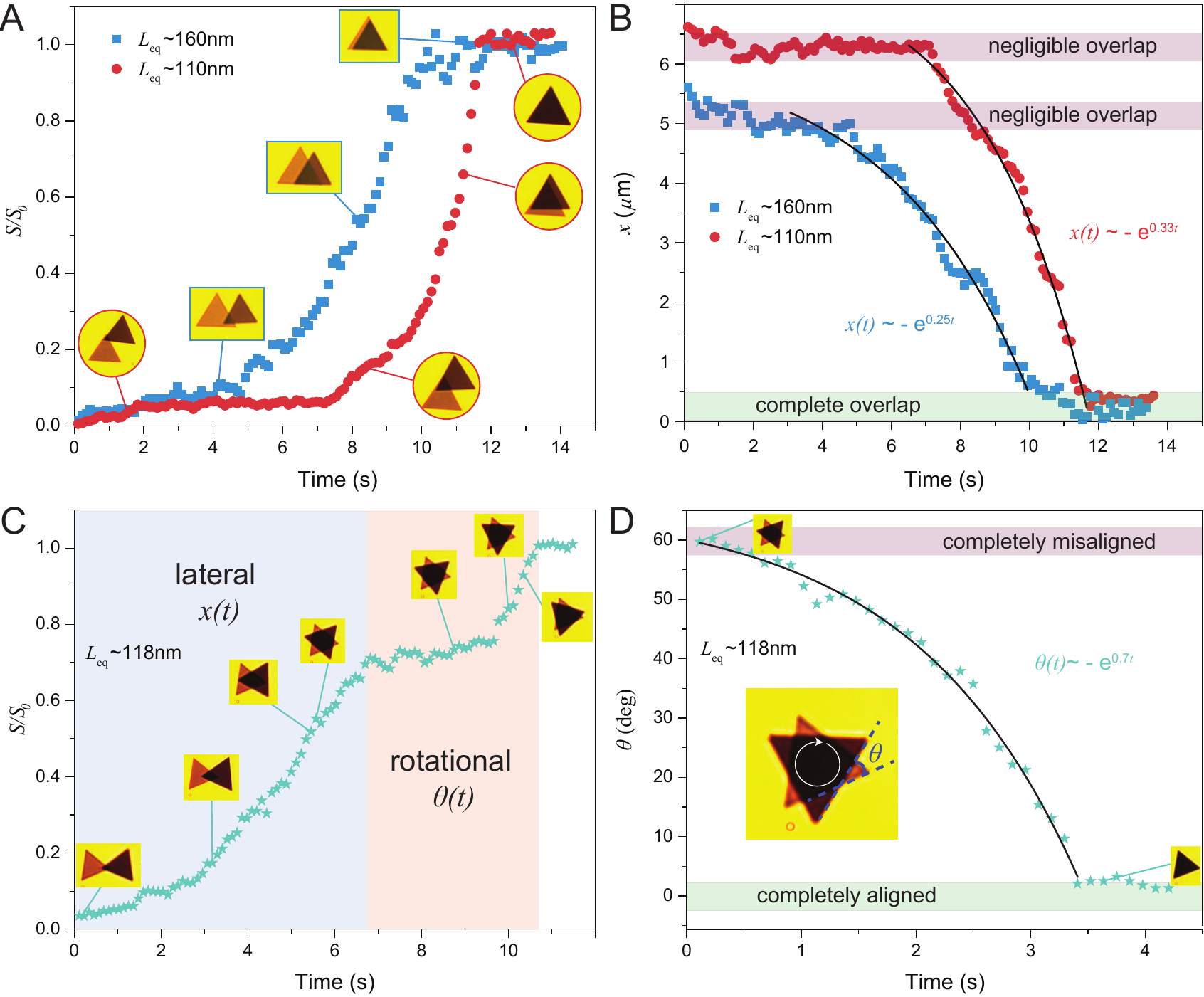}
\caption{\textbf{Formation of stable Casimir microcavities along lateral ($x$) and rotational ($\theta$) coordinates.}  \rev{(\textbf{A})} Evolution of the overlap area with time, $S(t)/S_0$, due to the Casimir self-alignment at two distinct equilibrium distances, $L_{\rm eq}$ = 160 nm and $L_{\rm eq}$ = 110 nm. The Au seed size is 7 \um in both cases. Insets depict exemplary video frames corresponding to data points marked on the $S(t)/S_0$ graph. \rev{(\textbf{B})} Evolution of lateral displacement between the centers of mass of the floating flake and the seed with time, $x(t)$, for the dimers shown in  \rev{(\textbf{A})}, eventually leading to a complete overlap, i.e. $x=0$, in both cases. Black curves are exponential fits of the experimental displacement data.  \rev{(\textbf{C})} Evolution of the overlap area between the seed and Au flake with time, $S(t)/S_0$, marking distinct phases of lateral and rotational motions for an $L_{\rm eq}$ = 118 nm dimer. The Au seed size is 7 $\mu$m. Insets depict exemplary video frames corresponding to data points marked on the $S(t)/S_0$ graph.  \rev{(\textbf{D})} Evolution of angle between the seed's and the floating flake's edges with time, $\theta(t)$, illustrating the self-alignment process. Black line marks an exponential fit of the experimental alignment data.}
\label{fig:Fig2}
\end{figure}

Stable dimers emerge as a result of the equilibrium between two opposing forces, the attractive Casimir force and the repulsive electrostatic force, occurring at a specific distance denoted as $L_{\rm eq}$. Furthermore, these dimers support optical Fabry-Pérot (FP) resonances that can be modified by controlling $L_{\rm eq}$. This is illustrated in Fig. \ref{fig:Fig1}\rev{E}, where variations in the reflected light's color are depicted for different $L_{\rm eq}$ values. The corresponding reflection spectra, as shown in Fig. \ref{fig:Fig1}\rev{F}, serve as the basis for the experimental determination of $L_{\rm eq}$ (see Methods). The variation of $L_{\rm eq}$ not only impacts the FP resonance but also substantially influences the Casimir potential, which scales approximately as $L_{\rm eq}^{-2.6}$ for 30 nm thick Au flakes \cite{Nature2021}. The manipulation of $L_{\rm eq}$ can be achieved by adjusting the total ion concentration in the solution, due to altering the Debye-Hückel screening length of electrostatic repulsion, described by $\kappa^{-1}$ (see Methods, Eq.~\ref{EqDebye}). Furthermore, the engineering of geometric patterns on the substrate allows to employ diverse sizes of triangular Au seeds and flakes (Fig. \ref{fig:Fig1}\rev{D}). This approach facilitates an investigation into the influence of overlap areas on the lateral Casimir force (Fig. \ref{fig:Fig1}\rev{B}). 

A specific example of Casimir self-alignment, depicted in Fig. \ref{fig:Fig1}\rev{G}, demonstrates the nearly isolated nature of two distinct motions: lateral and rotational (Supplementary Video 3), enabling their separate analysis. Throughout the self-assembly and self-alignment process, the floating Au flake initially exhibits predominantly lateral motion, moving towards the Au seed with minimal rotation. This lateral shift aims to increase the overlap area. Subsequently, in the second phase, the floating flake begins to rotate, ultimately achieving full overlap with the seed. It is intriguing to observe the individual contributions of these two motions to the overall increase in the overlap area and eventually forming a stable dimer (see Supplementary Videos 1-3).

In what follows, we analyze both, the dimer formation process and the stability of the self-assembled dimers to thermal fluctuations. Furthermore, by employing the seed concept, the system's behavior can be controlled by both $L_{\rm eq}$ and the total area of the Au seeds. The forthcoming sections provide a detailed exploration of how both of these parameters impact the Casimir potential, the dimer formation, the self-alignment process, and the robustness of these alignments to thermal fluctuations.

\vspace{0.5cm}
\noindent
{\large{\textbf{Accelerated dimer formation by lateral and rotational Casimir forces}}}
\vspace{0.5cm}

\noindent
Previously, we elucidated the advantages of employing the seed approach for achieving lateral motion and enhancing stability in lateral alignment with the corresponding seed. In this section, we additionally note that the lateral motion of the flake exhibits \emph{varying} speeds during the dimer formation with the seed. This phenomenon is especially evident when the floating flake approaches the seed with the same orientation. The lateral Casimir force draws the floating flake towards the Au seed, which causes a concurrent increase in the overlap area between the flake and the seed, $S(t)$. When the initial orientations of the triangles match, dimer formation occurs without any rotational motion, simplifying the observation of lateral motion acceleration as well as alterations in acceleration with increasing overlap area. This process is evidenced in Fig.~\ref{fig:Fig2}. Initially, when the overlap area is minimal, the flake's motion is slow and of predominantly Brownian nature. Over time, the seed exerts a robust attraction on the floating flake, eventually leading to full area overlap, $S_0$. The lateral and rotational motions stop when the two triangles achieve perfect overlap and alignment. Our data analysis involves extracting the overlap area of the seed and Au flake from each frame of the videos (Supplementary Videos 1 and 2) and normalizing it with the total area of the seed (or floating flake) -- $S(t)/S_0$, as depicted in Fig.~\ref{fig:Fig2}\rev{A}.

Since the separation distance is intricately linked to the Casimir potential, the acceleration of the Au flake exhibits a direct dependence on $L_{\rm eq}$. This is demonstrated in Fig. \ref{fig:Fig2}\rev{A}, where we present $S(t)/S_0$ for two distinct $L_{\rm eq}$ values. It is important to highlight the critical role played by $L_{\rm eq}$, given its strong influence on the lateral Casimir forces -- even slight alterations in $L_{\rm eq}$ yield a substantial impact on the acceleration dynamics of the Au flake. The inset pictures correspond to video frames depicting the exemplary data points. Furthermore, we calculate the position changes over time by tracing the displacement of the center of mass of the floating flake with respect to that of the seed, $x(t)$. 

Due to the viscous friction in the liquid and the small mass of the flake, the motion under the lateral Casimir force in our case occurs deeply in the overdamped regime \rev{($\Gamma_x>>\omega_x$, where $\Gamma_x$ is the oscillator's lateral damping constant normalized to the mass and $\omega_x$ is the oscillator's eigen frequency). In this regime, the role of the flake's mass is negligible, and the general solution of the equation of motion (see Supplementary Theory and Figs.~S4-S5) can be reduced to the sum of the fast $e^{-\Gamma_xt}$ (microsecond scale) and slow $-e^{\omega_x^2/\Gamma_xt}$ (second scale) decaying exponents with a negligible role of the former. Then, on a time scale of seconds, the lateral (and rotational) motion obeys the slow decaying exponential function $x(t)= \widetilde{a}\left(1-e^{\widetilde{\omega}(t-t_{\rm max})}\right)$, where $\widetilde{\omega}=\omega_x^2/\Gamma_x$ is a characteristic effective frequency and $\widetilde{a}$ is close to the triangle's edge size $a$ as long as $t_{\rm max}>>\widetilde{\omega}^{-1}$ (see Supplementary Theory). Notably, the \emph{non-harmonic} nature of the lateral overlap potential results in the \emph{inverse} curvature of this exponential function, while in a harmonic case, the decaying solution exhibits an ordinary positive curvature \rev{$x(t) \sim ae^{-\widetilde{\omega}t}$}.} The effective frequency $\widetilde{\omega} = |{U_0}|D_x\sqrt{3}\big/2k_{\rm B}T \sim$ 1 Hz is determined by the total potential per unit area $U_0$ (with Casimir and electrostatic contributions) and the lateral diffusion coefficient $D_x$. These contributions distinctly depend on $L_{\rm eq}$. Specifically, the absolute value of the total potential grows with a decrease in $L_{\rm eq}$ as follows: $|U_0| \sim |Ae^{-\kappa L_{\rm eq}}-BL_{\rm eq}^{-\alpha}|$, where $A$ and $B$ are distance-independent constants of the electrostatic and Casimir potentials, $\kappa^{-1}$ is the Debye–Hückel screening length, and $\alpha \approx 2.6$ is the Casimir potential power law for 30 nm thick Au flakes \cite{Nature2021}. On the contrary, the lateral diffusion coefficient decreases with a decrease in $L_{\rm eq}$ as $D_x\sim L_{\rm eq}^{\beta}$. However, according to previous hydrodynamic simulations, it exhibits a slower dependence with distance, with $\beta<1$~\cite{schmidt2023tunable}. Consequently, the effective frequency $\widetilde{\omega}$ is expected to increase with a decrease in $L_{\rm eq}$, so the lateral motion occurs faster when $L_{\rm eq}$ is smaller, as clearly illustrated by the exponential fits depicted in Fig.~\ref{fig:Fig2}\rev{B} \rev{(the fitting procedure is described in Methods)}. 

In Fig. \ref{fig:Fig2}\rev{C}, we employ the same method as in Fig. \ref{fig:Fig2}\rev{A} but for a different dimer configuration. As depicted in the plot, lateral motion plays a dominant role until $\approx$ 7 s mark, at which point the floating flake is found exactly on top of the seed but with an opposing orientation ($\theta \approx 60^{\circ}$). From that point on, the rotational motion becomes dominant, causing the floating flake to rotate around the vertical axis under the influence of the Casimir torque until it achieves full alignment (see Supplementary Video 3). Since the lateral motion becomes negligible after the first 7~s, we find it reasonable to analyze solely rotational motion by tracking changes in $\theta(t)$ over time, as depicted in Fig. \ref{fig:Fig2}\rev{D}. Similar to the lateral displacements, this rotational motion conforms to \rev{decaying exponential} function with an effective $\sim 1$ Hz frequency (see Supplementary Theory and \rev{Methods}), starting from a state of complete misalignment ($\theta \approx 60^{\circ}$) and culminating in a state of perfect alignment ($\theta = 0^{\circ}$). \rev{It is worth mentioning that generally the lateral and rotational motions are coupled and in most cases occur simultaneously, creating a system that is too complex to analyze (see Fig. S8). However, in a few particular cases (such as the ones illustrated in Figs.~\ref{fig:Fig1}-\ref{fig:Fig2}), we observed the two motions in a nearly isolated fashion, which allowed for their independent analysis.} 

Thus, our method results in the precise positioning (trapping) of floating Au flakes within three-dimensional space, offering control over their vertical, horizontal, and rotational orientations. This unique quantum trapping approach harnesses quantum vacuum fluctuations, rather than real fields (as in e.g. optical tweezing), for particle manipulation~\cite{ZhangScience2019,Nature2021}. Despite the inherent influence of thermal fluctuations, the trapping potential created by the Casimir effect exhibits remarkable stability, enabling controlled particle confinement. In the subsequent sections, we delve into an in-depth investigation of thermal fluctuations, specifically focusing on their magnitude and the resulting particle displacement, especially concerning rotational motion around the vertical axis.

\vspace{0.5cm}
\noindent
{\large{\textbf{Calculation of the lateral and rotational Casimir effect between two triangular flakes}}}
\vspace{0.5cm}

\noindent
To improve our understanding of the observed phenomenon and confirm its primary association with the Casimir effect, we conducted both analytical and numerical computations focusing on the in-plane motion of flat triangular Au flakes induced by vacuum fluctuations. The analytical calculations were performed using the Lifshitz formalism~\cite{lifshitz1956,Lifshitz-Dzyaloshinskii-Pitaevski1961} and involved several critical approximations to model the underlying dynamics.

First, the Casimir-Lifshitz potential $U_{\rm Lif}$ per unit area for the system comprising two identical infinite Au planar mirrors immersed in an aqueous solution and separated by a gap of thickness $L$ was calculated using the following expression:
\begin{equation}
        U_{\rm Lif}(L)=\frac{\hbar c}{4\pi^2}\int_{0}^{\infty }\rm{d}\xi \int_{0}^{\infty} \mathit{k_\parallel}\rm{d}\mathit{k_\parallel} \sum_{p,s}
        \ln{\left(1-\mathit{r}_{\rm p,s}^2(\mathit{i\xi},\mathit{k_\parallel},\mathit{d}_{\rm Au})\mathit{e}^{-2\mathit{L}\sqrt{\mathit{k_\parallel}^2+\xi^2\varepsilon_{\rm H_2O}(\mathit{i\xi})}}\right)},
        \label{EqLifshitz} 
\end{equation}
where the integration was performed over the imaginary frequencies $\xi=i\omega/c$, normalized by the speed of light $c$, and the wave vector components along the mirrors $k_\parallel$. The summation involved both p- and s-polarizations. Here, $r_{\rm p,s}(i\xi,k_\parallel,d_{\rm Au})$ are the Fresnel reflection coefficients for the gold plates evaluated at various wave vectors and mirror thicknesses $d_{\rm Au}$. The dielectric functions of gold $\varepsilon_{\rm Au}(i\xi)$ and water $\varepsilon_{\rm H_2O}(i\xi)$ were evaluated at the imaginary frequencies.

Second, to account for the shape and size of the floating flakes and seeds, we employed the overlap approximation. This approximation assumes that the Casimir-Lifshitz potential is proportional to the overlap area between the flake and the seed while neglecting any edge effects. In this approach, the lateral potential can be expressed as follows:
\begin{equation}
E_{\rm lat}(x,y,\theta)\approx S(x,y,\theta)U_{\rm Lif}(L_{\rm eq}),
\end{equation}
where $x,y$ are the in-plane displacements, $\theta$ is the in-plane rotation angle, $S(x,y,\theta)$ is the overlap area between the flake and the seed, and $U_{\rm Lif}$ is the Casimir-Lifshitz potential per unit area evaluated at $L_{\rm eq}$.

Third, we assume that the lateral displacements (along $x$ or $y$) and rotations (around $z$ by angle $\theta$) are \emph{independent} of each other, allowing for factorization. \rev{Certainly, this assumption is valid only in some particular cases, e.g. the ones analyzed in Figs.~\ref{fig:Fig1}-\ref{fig:Fig2}. In general, lateral and rotational motions are not independent, and require more involved numerical simulations, as discussed below.} \rev{In the independent scenario, however}, we arrive at simplified forms for the potentials: $E_x(x)\approx S(x)U_{\rm Lif}(L_{\rm eq})$ for lateral displacement (along $x$) \rev{of perfectly aligned triangles} and $E_\theta(\theta)\approx S(\theta)U_{\rm Lif}(L_{\rm eq})$ for rotation \rev{of triangles with the same position of the center of mass}. In our experiments, we focused on equilateral triangles, where the corresponding overlap areas are expressed as follows: 
\begin{equation}
S(x)\rev{\Big|_{\theta=0}}=S_0\left(\frac{a-x}{a}\right )^2, \;\;\;
S(\theta)\rev{\Big|_{x=0}}=S_0\frac{\rm{tg}(\pi/3-\theta/2)-\rm{tg}(\theta/2)}{\sqrt{3}},
\end{equation}
where $S_0$ is the area of an equilateral triangle with the edge $a$. Lateral Casimir forces and Casimir torques can be obtained by differentiating $E_x$ and $E_{\theta}$, correspondingly, along their respective motions (see Supplementary Theory).

\begin{figure}[h]
\includegraphics[width=1.\textwidth]{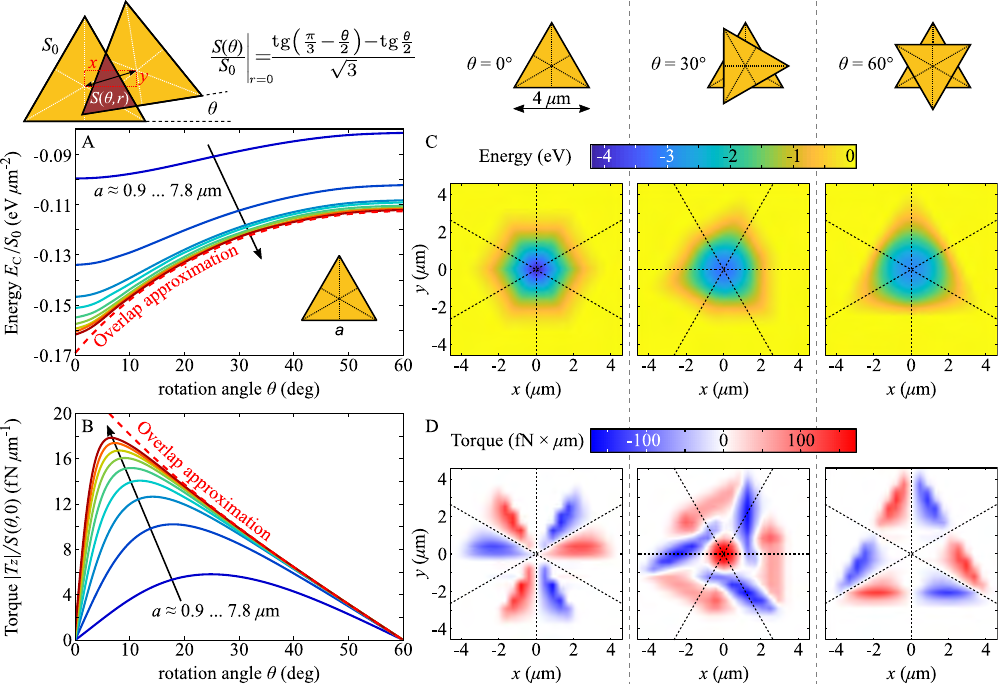}
\caption{\textbf{Calculation of the lateral Casimir interactions of two gold equilateral triangles.} \rev{The triangles have} edge $a$, thickness \unit[40]{nm} and surface area $S_{0}$, \rev{are} vertically trapped at $L_{\rm eq}$ = 200 nm and immersed in aqueous solution, for the relative rotation by angle $\theta$ and/or translation by $r=(x, y)$.  \rev{(\textbf{A})} Casimir energy \emph{vs.} the rotation angle of one triangle around the center of mass. The energy normalized to triangle surface area $S_{0}$ increases with size until the limit determined by the overlap area, $S(\theta, r=0)$, of two mutually rotated triangles. \rev{(\textbf{B})} Casimir torque $T_z$ around an axis normal to the triangle surface and going through the center of mass normalized to the overlap area $S(\theta,0)$ \emph{vs.} $\theta$, exhibiting an asymmetric increase towards small angles. 
\rev{(\textbf{C, D})} Casimir energy and torque for two triangles ($a=4$~$\mu$m) as a function of both rotation around the center of mass ($\theta = 0^\circ,\, 30^\circ,\, 60^\circ$) and translation $(x, y)$. The dashed lines mark the symmetry axes of the rotated triangle.}
\label{fig:Fig3}
\end{figure}

To assess the significance of the edge effects, we conducted a comparative analysis between the analytical calculations within the overlap approximation and numerical simulations using SCUFF-EM~\cite{Reid,SCUFF}. Specifically, we simulated the relative rotation and in-plane translation of two vertically trapped Au equilateral triangles immersed in an aqueous solution, where the Casimir and electrostatic forces at $L_{\rm eq}$ are balanced. Figure~\ref{fig:Fig3} presents the simulation results alongside analytical calculations. Due to the symmetry of the equilateral triangles, we considered rotation angles ranging from $0^\circ$ to $60^\circ$. Examining the rotational dependence of the Casimir energy normalized to the corresponding triangle surface area at zero $x$ and $y$ displacements, it is evident that the numerical curve approaches the analytical approximation as the triangle size increases. For triangles with $a \sim 8$ \um, the overlap approximation performs exceptionally well for $\theta>10^\circ$ (Fig.~\ref{fig:Fig3}\rev{A}). As mentioned earlier and detailed in Supplementary Theory, the lateral Casimir potential has a non-harmonic shape in the overlap approximation. However, at small angles, the edge effects captured by SCUFF-EM simulations become prominent, resulting in a quasi-harmonic potential, as shown in Fig.~\ref{fig:Fig3}\rev{A (see also Supplementary Theory and Figs. S6-S7)}. This effect is even more pronounced for the Casimir torque, normalized to the overlap area $S(\theta,0)$ (Fig.~\ref{fig:Fig3}\rev{B}). Here, the overlap approximation performs well at $\sim \theta>20^\circ$. However, at small angles, the edge effects become considerable. Importantly, the torque cannot surpass the limit set by the overlap approximation at any angle. Furthermore, the torque reaches its maximum at an angle where an optimal balance between edge effects and increased overlap area is attained. With increasing triangle size, the role of edge effects diminishes, causing the optimal angle to decrease, and eventually approach zero for infinite triangles. Notably, unlike the anisotropic finite-sized systems considered by Antezza \textit{et al.}~\cite{AntezzaTorque2020}, we observe high accuracy of the overlap approximation for large triangles owing to their isotropy.

For non-zero in-plane displacements of the triangles, the discussed phenomena become even more intricate. Figures~\ref{fig:Fig3}\rev{C, D} show the combined influence of translation and rotation on the Casimir energy and torque, respectively. The energy's dependence on displacements consistently reaches a minimum in the center at $x=y=0$ (Fig.~\ref{fig:Fig3}\rev{C}), corresponding to the stable configuration of the system and the lateral Casimir trapping. As rotation is introduced ($\theta \neq 0$), the energy colormap undergoes a shape transition from hexagon to triangle forms, although this does not lead to qualitative changes. The role of rotation is evident in the torque colormaps (Fig.~\ref{fig:Fig3}\rev{D}). For a non-zero rotation angle, in the region of zero displacements, a pronounced spot of maximum torque emerges, aligning with the earlier discussed scenario of rotation at zero displacements. Remarkably, even at zero rotation, the interplay between the system's geometry and relative displacements may lead to non-vanishing torques. This effect disappears on the symmetry axes of the rotated triangle, having different signs on opposite sides of these axes. Additionally, at $\theta=60^\circ$, the rotated system becomes so symmetric that the torque is entirely compensated in the large area around zero displacements. For all intermediate angles (additional data shown in Supplementary Material) between $0^\circ$ and $60^\circ$, a mixed pattern emerges, with maximum torque at the center, surrounded by areas of effective torque induced by displacements and rotations. These areas deviate from the symmetry axes of the rotated triangle but maintain alternating torque signs around the central maximum. Consequently, as the flake moves with simultaneous displacement and rotation, additional torque in the opposing direction may arise, resulting in a crawling-like motion.

\vspace{0.5cm}
\noindent
{\large{\textbf{}}}
\vspace{0.5cm}

\noindent
The self-alignment experiments are conducted in the $L_{\rm eq}$ range spanning $\sim$ 100 -- 200 nm, where conditions allow establishing a stable equilibrium and a sufficiently deep trapping potential. In each self-alignment measurement, $L_{\rm eq}$ is precisely determined by assessing the reflection spectrum of the stable FP microcavity, Fig. \ref{fig:Fig1}\rev{F}, and subsequently fitting the spectrum using the transfer-matrix method (see Methods).

To investigate the impact of $L_{\rm eq}$ on self-alignment, we maintain a fixed seed size, $a = 4$ $\mu$m, while choosing a floating flake that matches this seed size. Optical tweezers are employed to bring the flake into close proximity with the seed, and then allow it to freely diffuse until the Casimir potential ensures the formation of a stable dimer. Subsequently, we track the dynamics of the angle between the triangles over time, $\theta(t)$. When $\theta=0$, the floating Au flake perfectly aligns with the seed, and the Casimir potential is minimized, thus corresponding to the equilibrium position, as illustrated in Fig. \ref{fig:Fig3}\rev{A}.

\begin{figure}
\centering
\includegraphics[width=0.7\textwidth]{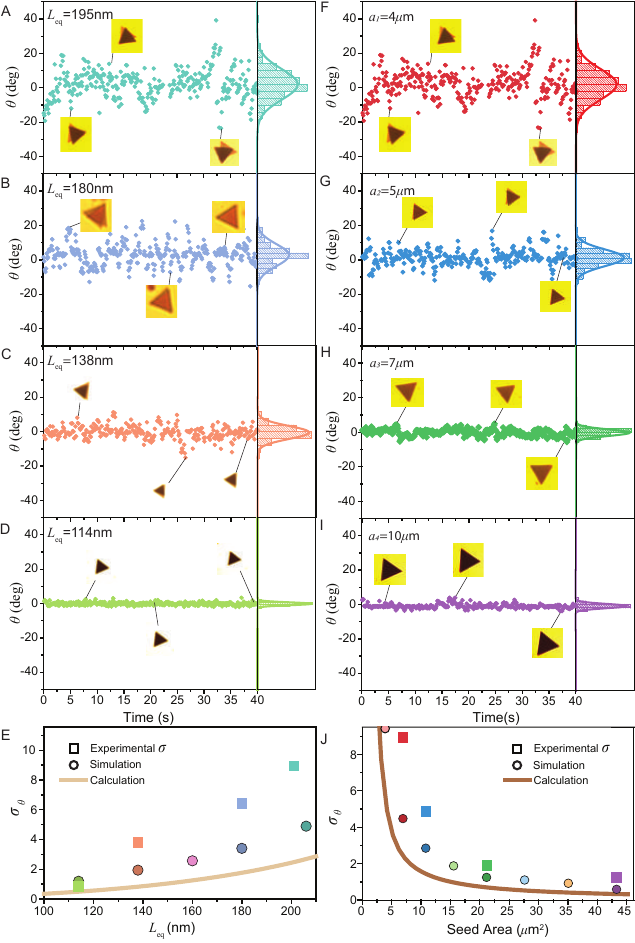}
\caption{\textbf{Impact of equilibrium distance and flake's area on stability of self-alignment to thermal fluctuations.} Variation of angle $\theta(t)$ in time for stable dimers of $a=4$ $\mu$m seed and a floating flake of similar size for $L_{\rm eq}=195$ nm \rev{(\textbf{A})}, $L_{\rm eq}=180$ nm \rev{(\textbf{B})}, $L_{\rm eq}=138$ nm \rev{(\textbf{C})}, and $L_{\rm eq}=114$ nm \rev{(\textbf{D})}. The distributions of $\theta$ are shown on the right of the corresponding time traces and are fitted to a normal distribution. Insets depict exemplary video frames corresponding to data points on the graph. \rev{(\textbf{E})} Standard deviation ($\sigma_{\theta}$) $vs.$ $L_{\rm eq}$. Variation of angle $\theta(t)$ in time for stable dimers at $L_{\rm eq} \approx 195$ nm and seeds with lateral sizes of 4 $\mu$m \rev{(\textbf{F})}, 5 $\mu$m \rev{(\textbf{G})}, 7 $\mu$m \rev{(\textbf{H})}, and 10 $\mu$m \rev{(\textbf{I})}. In all cases, the sizes of floating flakes were approximately the same as the corresponding seeds. The normal distribution of the $\theta$ is shown on the right side of each panel. Insets images indicate the video frames to the corresponding data points on the graph. \rev{(\textbf{J})} Standard deviation, $\sigma_{\theta}$, for various Au seed areas. In \rev{(\textbf{E})} and \rev{(\textbf{J})} the simulation points are obtained using SCUFF-EM and the calculation curves -- with the Lifshitz formalism in the overlap approximation.}
\label{fig:Fig4}
\end{figure}

Figures~\ref{fig:Fig4}\rev{A-D} illustrate the variations in $\theta(t)$ for four distinct $L_{\rm eq}$ values: \rev{195} nm, 180 nm, 138 nm, and 114 nm, respectively, as a result of thermal fluctuations. Specifically, in Fig.~\ref{fig:Fig4}\rev{A}, the fluctuations in $\theta$ are the strongest among situations presented in Fig.~\ref{fig:Fig4}. This is attributed to the fact that at this $L_{\rm eq}$, the thermal energy ($k_{\rm B}T$) approaches the depth of the trapping potential, leading to more pronounced deviations in $\theta(t)$ from its equilibrium position due to thermal fluctuations. As a result, the floating flake exhibits fluctuations of up to $\pm$20$^{\circ}$ at different points in time (see $\theta$ distributions and standard deviations to the right of the corresponding $\theta(t)$ plots). 

Importantly, the fluctuations of $\theta$ decrease upon reduction in $L_{\rm eq}$. To demonstrate that, we fit the experimental histograms of $\theta(t)$ to the normal distribution functions, and find that the variances, denoted as $\sigma_{\theta}$, grow with $L_{\rm eq}$, as depicted in Fig.~\ref{fig:Fig4}\rev{E}. The reason for this lies in the steep dependence of the trapping Casimir potential on the distance, $U_{\rm C}\propto L_{\rm eq}^{-2.6}$ \cite{Nature2021} as discussed above. Furthermore, we determine $\sigma_{\theta}$ theoretically by substituting the analytical Casimir-Lifshitz potential from Eq.~(\ref{EqLifshitz}) into the equation for the mean square displacement using Gibbs formalism (see Methods, Eq.~\ref{Eqstd}). Combined with the overlap approximation and the Casimir potential from SCUFF-EM simulations, we arrive at results that are qualitatively similar to experimental observations. Our theoretical results are especially close to the experimental findings at small $L_{\rm eq}$, however, they underestimate $\sigma_{\theta}$ approximately 2-fold (for SCUFF-EM simulations) at large $L_{\rm eq}$. This discrepancy can be attributed to electrostatics, which was not included in our calculations and modeling, given its relatively weak contribution compared to Casimir forces. Indeed, the relative contribution of the electrostatic part to the total potential increases with $L_{\rm eq}$, contributing to the growing mismatch at large separations. Furthermore, the difference between the analytical and numerical results reflects the significance of the edge effects, especially at small deviations from equilibrium. Additionally, the accuracy of the overlap approximation not only increases with the size of the flakes ($a$), as we discussed earlier, but also with a decrease in $L_{\rm eq}$. In other words, the analytical approximation's accuracy increases with the ratio $a/L_{\rm eq}$. Finally, it is important to note that at small displacements ($x$ and $y$), the Casimir torques (and hence $\theta(t)$) are almost independent of displacements (see Fig.~\ref{fig:Fig3}\rev{D}), justifying our analysis.

The magnitude of the trapping potential scales with the area overlap of the seed and the floating Au flake. This is a fundamental reason for the rotational self-alignment observation in the seed configuration -- Au flakes in the aqueous solution strive to maximize their overlap areas with the seeds due to strong Casimir attraction. As previously, we preselect the flakes such that they match the shape and size of the corresponding seed, which simplifies experiments and allows for precise tracking of the rotational motion of the floating flake $\theta(t)$ (see Supplementary Video 4).

In the subsequent study, we probe the stability of the trapping potential to thermal fluctuations for a fixed $L_{\rm eq}$ $\approx$ 195 nm, as a function of the total area of the seed, $S_0$ (Fig.~ \ref{fig:Fig1}\rev{B}). We specifically investigate four different seed sizes in the range between 4 and 10 $\mu$m, depicted in Fig.~\ref{fig:Fig4}\rev{F-I}. As expected from the Casimir energy calculations (see Fig.~\ref{fig:Fig3}\rev{A}), increasing the area of Au seeds increases the Casimir energy (and the depth of the trapping potential), which results in more stable microcavities. 

The stability dependence on the overlap area is shown in Fig.~\ref{fig:Fig4}\rev{F-I} by plotting the dynamics of $\theta(t)$ caused by thermal fluctuations for $a=$ 4, 5, 7, and 10 \um triangle seed sizes. The standard deviations, $\sigma_{\theta}$, obtained from fits of $\theta(t)$ histograms to the normal distribution function, indicate that the stability strongly depends on the area overlap. In particular, the biggest flakes display the most pronounced stability -- i.e. the stability is inversely proportional to the flake area, or more accurately, $\sigma_{\theta} \propto S_0^{-1.2}$ \rev{(Fig.~\ref{fig:Fig4}J)}. This aligns with the analytical calculations and numerical simulations discussed in the previous paragraphs. However, in this case, the difference between the calculations and measurements remains consistent across different overlap areas. This further validates that the discrepancy is indeed caused by the contribution of the electrostatic potential. These results demonstrate that control of lateral stability of the FP cavity can be achieved by selecting various seed sizes while maintaining the same separation distance (see Supplementary Video 5).

\rev{Finally, we note that the stability of the self-alignment process is not notably  impacted by the Au flake (or seed) thickness within the range of 20 -- 35 nm used in our experiments. The resulting variance of the Casimir potential is less than 5\% (see Fig.~S9, where we present a theoretical calculation based on Lifshitz formalism).}

In conclusion, vertically and laterally stable self-assembled and rotationally self-aligned Casimir microcavities can be formed by equilibrating two opposing forces: attractive Casimir and repulsive electrostatic. The so-formed microcavities exhibit optical Fabry-Pérot resonances in the visible spectral range, observed as pronounced reflection colors. Control of the self-alignment of the cavity is demonstrated by using triangular templated substrates and is studied upon variation of two independent parameters: (\emph{i}) the equilibrium distance, $L_{\rm eq}$, between the floating flake and the templated substrate, and (\emph{ii}) the flake area, $S_0$, by varying the seed and the floating flake sizes. These two parameters are straightforward to control. Therefore, our method offers flexibility to achieve the desired conditions for liquid-phase Casimir torque experiments. Furthermore, we investigated the stability of self-alignment to thermal fluctuations, which was assessed by examining $\theta(t)$ as a function of $L_{\rm eq}$ and $S_0$. We find that our experimental observations are in good agreement with Casimir-Lifshitz theory and with SCUFF-EM numerical modeling. This work presents a new self-assembly and self-alignment platform based on quantum trapping and templated substrates that is suitable for Casimir torque experiments. Furthermore, the presented FP microcavities are stable at room temperature for as long as they have been monitored, which offers a possibility of their future use in nanomachinery~\cite{ZhangScience2019}, self-assembly~\cite{batistareview2015}, optomechanics~\cite{eichenfield2009picogram}, polaritonic chemistry~\cite{thomas2019tilting}, and other potential cavity-inspired applications~ \cite{torma2014strong,baranov2017novel,garcia2021manipulating}.   

\vspace{0.5cm}
\noindent
{\large{\textbf{Author information}}}
\vspace{0.2cm}

\noindent
\textbf{Corresponding author.} Timur O. Shegai, Email: timurs@chalmers.se\\
\textbf{Notes.} The authors declare no competing financial interest.

\vspace{0.5cm}
\noindent
{\Large{\textbf{Methods}}}
\vspace{0.2cm}

\noindent
{\large{\textbf{Seed Fabrication}}}
\vspace{0.2cm}

\noindent
All seed samples were prepared on thin (170 $\mu$m) microscope glass coverslips. The glass coverslips were cleaned in acetone, 2-propanol, and water at 50~°C in an ultrasonicator for 15 min for each solvent. Subsequently, the coverslips were dried using compressed nitrogen, followed by oxygen plasma cleaning. Triangular Au seed arrays with various edge sizes were fabricated using standard electron-beam lithography (Raith EBPG 5200). To provide the adhesion of the Au to the glass substrate, first, a 2~nm Cr layer was evaporated. Subsequently, a 20~nm Au layer was evaporated. Both layers were evaporated using the Kurt J. Lesker PVD 225 tool. 

\vspace{0.5cm}
\noindent
{\large{\textbf{Gold Flake Synthesis and KBr Preparation}}}
\vspace{0.2cm}

\noindent
Single crystal Au flakes were synthesized using an aniline-assisted method in ethylene glycol (EG) \cite{anilinesynt2006}. This synthesis method is preferred due to its ability to produce colloidal microflakes with a large aspect ratio between their lateral size and thickness. Briefly, the synthesis protocol includes the following steps. First, a 0.72 mM HAuCl$_4$$\cdot$3H$_2$O solution in 50~ml EG is prepared in a glass bottle and heated to 95~°C in a water bath for 20 min. Second, a 0.1~M aniline solution in EG is added to the heated solution under mild stirring until the molar ratio reaches 2:1 of aniline to gold. In order to reach this molar ratio, a 0.72~ml of 0.1~M aniline is added in our case, since our initial solution contained a 50~ml of 0.72~mM HAuCl$_4$$\cdot$3H$_2$O. When the aniline is mixed homogeneously, the reaction is kept undisturbed in a water bath at 95~°C for 3~h. This synthesis protocol yields Au flakes with a high aspect ratio, with flake thicknesses ranging from 20 to 35 nm and lateral dimensions varying from 3 to 15 $\mu$m. After successful synthesis, the solution contains a large number of precipitated Au flakes on the bottom or walls of the bottle. Additionally, the synthesis yields a large amount of byproduct -- quasi-spherical Au particles in the solution. Therefore, it is important to remove the spherical particles as well as to replace the EG with cetyltrimethylammonium bromide (CTAB) aqueous solution. This replacement requires several steps of washing. Specifically, after the Au flakes sediment, half of the volume of the supernatant is removed and replaced with the same volume of pure 20~mM CTAB aqueous solution. The new mixture is shaken to disperse the CTAB and form a double layer on the surface of the Au flakes. This step is repeated several times until most of the spherical particles are removed and EG is replaced with the 20~mM CTAB solution. Once the replacement of EG is completed, the CTAB concentration in the solution is further diluted to the 1~mM level. The final synthesis batch is stored at this CTAB concentration for subsequent use. In self-alignment experiments, the CTAB concentration was further diluted to 0.1~mM and the final ionic environment was adjusted with an aqueous solution of potassium bromide (KBr). To do that, KBr salt is dissolved in deionized water in the desired concentration and subsequently mixed with the 0.1~mM CTAB. The total ion concentration in solution is a parameter that is directly connected to the Debye-Hückel screening length, defined as: 
\begin{equation}
\kappa^{-1}=\sqrt{\frac{\varepsilon_{H_2O}(0)\varepsilon_0 k_{\rm B} T}{C_{tot} q_0^2 z^2}} 
\label{EqDebye} 
\end{equation}

where $\varepsilon_{H_2O}(0)$ is the static permittivity of water, $\varepsilon_0$ is the vacuum permittivity, $q_0$ is the elementary charge,  $C_{tot}$ is the total ion concentration with valence $z$ (in this case ions are CTA$^+$ and Br$^-$ from CTAB and K$^+$ and Br$^-$ from salt, therefore, $z=1$ in all cases).

\vspace{0.5cm}
\noindent
{\large{\textbf{Optical measurements}}}
\vspace{0.2cm}

\noindent
All reflectivity and self-alignment measurements were performed using an inverted microscope (Nikon Eclipse TE2000-E) equipped with an oil immersion 100$\times$ objective with an adjustable numerical aperture (NA = 0.5 -- 1.3). All reflection spectra are taken at quasi-normal incidence using NA = 0.5 and a halogen light source, the reflected spectrum is directed to the fiber-coupled spectrometer (Andor, Shamrock 500i), equipped with a CCD camera (Andor, Newton 920). The equilibrium distance, $L_{\rm eq}$, was assessed by fitting the experimentally measured reflectivity spectra with the transfer-matrix method \rev{(see Figs. S1-S3)}. 

For the selection of the Au flakes with the correct size and shape, the optical tweezers method was employed. To trap the Au flakes, a $\lambda = 447$~nm continuous wave laser with a power of 9~mW was focused on the desired flake through a 100$\times$ objective (NA = 1.3). All self-alignment data is collected through recording videos using Thorlabs DCC1645C-HQ camera and each video frame is analyzed using MATLAB, allowing to extract the dynamics of angle, $\theta(t)$, for subsequent analysis of the Casimir self-alignment process.

\vspace{0.5cm}
\noindent
{\large{\textbf{Fitting procedure}}}
\vspace{0.2cm}

\noindent
\rev{To fit the experimental observations of lateral and rotational microcavity formation observed in Fig.~\ref{fig:Fig2}, we used the decaying exponential solution described in detail in the Supplementary Theory. The expression for the lateral motion reads: $x(t)=\widetilde{a}\left(1-e^{(t-t_{\rm max})\widetilde{\omega}}\right)$, 
where $\widetilde{a}=a (1 - e^{-t_{\rm max}\widetilde{\omega}} )^{-1}$ and $a$ is the triangle's edge size. The fitting parameters for the lateral motion shown in Fig.~\ref{fig:Fig2}B are as follows: $\widetilde{a}_{160}=6.2$ $\mu$m, $\widetilde{\omega}_{160}=0.25$ s$^{-1}$, $t_{\rm max}=10.3$ s for $L_{\rm eq}$ = 160 nm cavity; and $\widetilde{a}_{110}=7.7$ $\mu$m, $\widetilde{\omega}_{110}=0.33$ s$^{-1}$, $t_{\rm max}=11.7$ s, for $L_{\rm eq}$ = 110 nm cavity, respectively. For rotational motion, shown in Fig.~\ref{fig:Fig2}D, we similarly use $\theta(t)=\widetilde{\theta}_0\left(1-e^{(t-t_{\rm max})\widetilde{\omega}}\right)$, 
where $\widetilde{\theta}_0=\theta_0(1 - e^{-t_{\rm max}\widetilde{\omega}})^{-1}$ and $\theta_0 = 60^{\circ}$. The extracted parameters are: $\widetilde{\theta}_0 = 66^{\circ}$, $\widetilde{\omega}=0.7$ s$^{-1}$, $t_{\rm max}=3.5$ s.}

\vspace{0.5cm}
\noindent
{\large{\textbf{Calculations of the angular spread}}}
\vspace{0.2cm}

\noindent
To theoretically determine the standard deviation of angle $\theta$ describing rotations under the lateral Casimir potential $E(\theta)$ and subjected to thermal fluctuations at temperature $T$, we utilized the expression for the mean squared displacement in Gibbs formalism: 
\begin{equation}    \sigma_\theta=\sqrt{\left\langle\theta^2\right\rangle}=\sqrt{\int_{-60^\circ}^{60^\circ}\theta^2 e^{-E(\theta)/k_{\rm B}T}\rm{d}\theta\bigg/\int_{-60^\circ}^{60^\circ} \mathit{e}^{-\mathit{E(\theta)/k_{\rm B}T}}\rm{d}\theta},
\label{Eqstd}     
\end{equation}

where we account for the symmetry of the potential, implying $\left\langle\theta\right\rangle=0$. Substituting the analytical Casimir-Lifshitz potential from Eq.~(\ref{EqLifshitz}), combined with the overlap approximation $E(\theta) \approx S(\theta)U_{\rm Lif}(L_{\rm eq})$ and the Casimir potential from the SCUFF-EM simulations, we obtained the final results denoted as calculations and simulations, respectively, in Fig.~\ref{fig:Fig4}.

\vspace{0.5cm}
\noindent
{\large{\textbf{Acknowledgements}}}
\vspace{0.2cm}

\noindent
\rev{\textbf{Funding:} The authors acknowledge financial support from the Swedish Research Council (VR Miljö project, grant No: 2016-06059, VR project, grant No: 2017-04545, and VR project, grant No: 2022-03347), the Knut and Alice Wallenberg Foundation (grant No: 2019.0140), Chalmers Area of Advance Nano, 2D-TECH VINNOVA competence center (Ref. 2019-00068), and Olle Engkvists foundation (grant No: 211-0063). T.J.A. acknowledges support from the Polish National Science Center via the project 2019/34/E/ST3/00359. The computations were enabled by resources provided by the Swedish National Infrastructure for Computing (SNIC) at PDC and by the Interdisciplinary Centre for Mathematical and Computational Modelling at the University of Warsaw (Grant \#G55-6). \textbf{Author Contribution:} B.K., O.V.K., and T.O.S. conceived the idea and planned the project. B.K. performed all experiments, analyzed the data, and performed TMM fittings. A.C., A.Y.P., A.V.A. contributed to the nanofabrication of the seed structures. O.V.K. and T.J.A. performed the analytical and numerical calculations. B.K., O.V.K., T.J.A., and T.O.S. wrote the manuscript with the help of all authors. \textbf{Competing interests:} The authors declare that they have no competing interests. \textbf{Data and materials availability:} All data needed to evaluate the conclusions in the paper are present in the paper and/or the Supplementary Materials.}



\newpage
\setcounter{page}{1}

\setcounter{figure}{0}
\setcounter{equation}{0}
\newcounter{note}
  
\renewcommand{\theequation}{S\arabic{equation}}
\renewcommand{\thesection}{S\arabic{section}}
\renewcommand{\thefigure}{S\arabic{figure}}
\renewcommand{\thetable}{S\arabic{table}}
\renewcommand{\thenote}{S\arabic{note}}
\newcommand{\safece}[1]{\texorpdfstring{\ce{#1}}{#1}}
\newcolumntype{C}[1]{>{\centering\arraybackslash}p{#1}}

\makeatletter
\newcommand{\manuallabel}[2]{\def\@currentlabel{#2}\label{#1}}
\makeatother

\newcommand{\notetitlelabel}[2]{
\phantomsection
\refstepcounter{note}
\addcontentsline{toc}{subsection}{\ref{#1}. #2}
\manuallabel{#1}{\thenote}
\textbf{Supporting Note \thenote: #2.}
}

\begin{center}

{\large %
Supplementary Materials
}
\vspace{2ex}

{\Large\bf %
Quantum trapping and rotational self-alignment in triangular Casimir microcavities
}
\vspace{3ex}

Betül Küçüköz,$^{1,\ast}$ Oleg V. Kotov,$^{1,\ast}$ Adriana Canales,$^{1,\ast}$ Alexander Yu. Polyakov,$^{1}$ Abhay V. Agrawal,$^{1}$ Tomasz J.\ Antosiewicz,$^{2,1}$ and Timur O. Shegai$^{1,\dagger}$

\vspace{1ex}

\emph{
$^{1}$Department of Physics, Chalmers University of Technology, 412 96, Gothenburg, Sweden \\
$^{2}$Faculty of Physics, University of Warsaw, Pasteura 5, 02-093 Warsaw, Poland \\
$^{\ast}$These authors contributed equally to this work. \\
$^{\ddagger}$email: \href{mailto:timurs@chalmers.se}{timurs@chalmers.se} \\
}

\end{center}

\vspace{10mm}

\tableofcontents

\clearpage

\section{Supplementary Figures}

\FloatBarrier

\begin{figure}[h]
\includegraphics[width=1\textwidth]{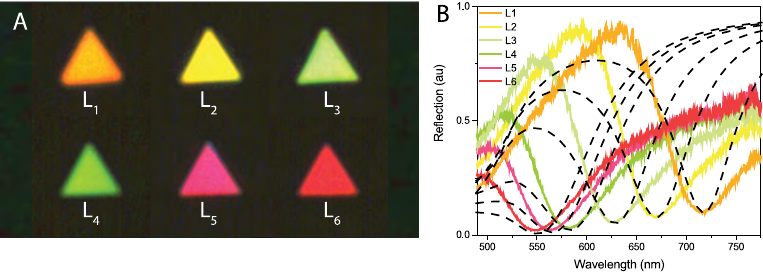}
\caption{\textbf{Stable Casimir microcavities with different \Leq values.} (\textbf{A}) True-color reflectivity images of self-aligned microcavities using 5 \um triangular seeds with various \Leq values measured using an inverted microscope. The same image is shown in Fig.~1e of the main text. (\textbf{B}) Reflection spectra of the microcavities from (\textbf{A}) together with corresponding TMM fits (black dashed lines). \Leq values are obtained as $L_1$ = 199 nm, $L_2$ = 179 nm, $L_3$ = 161 nm, $L_4$ = 140 nm, $L_5$ = 131 nm, $L_6$ = 123 nm.}
\label{Fig.S1}
\end{figure}

\begin{figure}[h]
\includegraphics[width=1\textwidth]{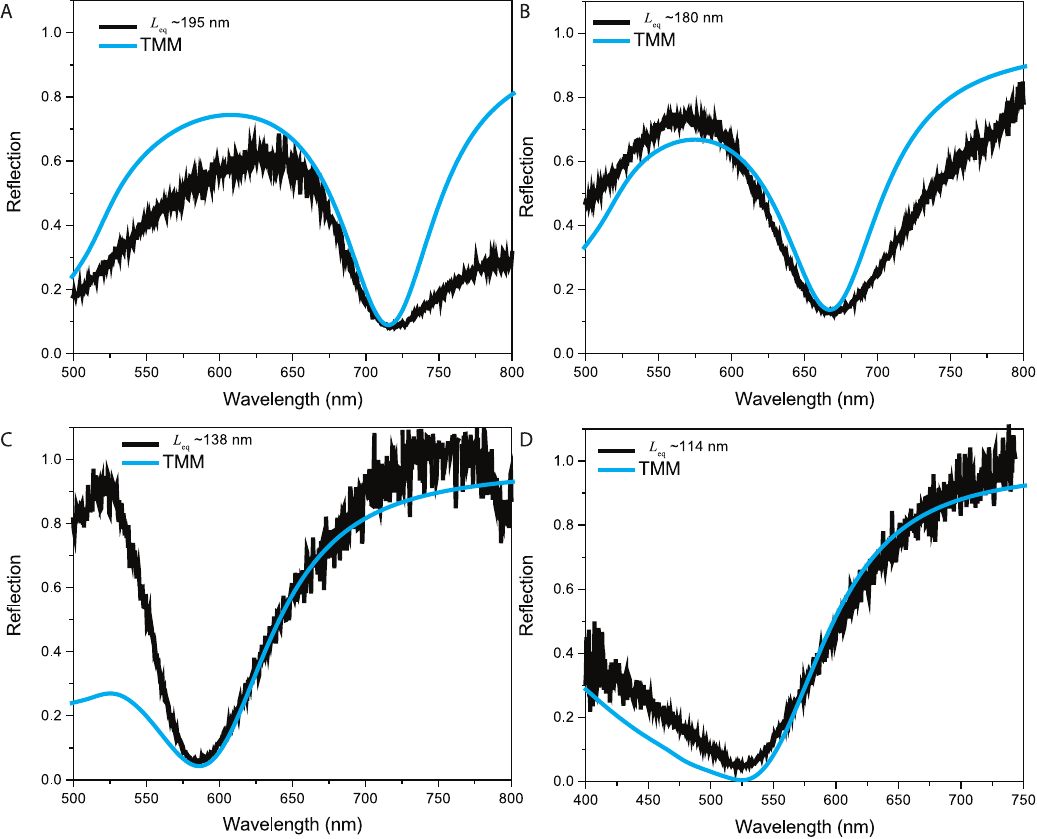}
\caption{\textbf{TMM fitting of the reflecitivty measurements from 4 \um flake with different \Leq values.} (\textbf{A-D}) Experimental reflection spectra and their transfer-matrix method (TMM) fittings for various $L_{\rm eq}$. This procedure allows to determine \Leq distance for Fabry-Pérot microcavities shown in Fig.~4A-D of the main text.}
\label{Fig.S2}
\end{figure}

\begin{figure}[h]
\includegraphics[width=1\textwidth]{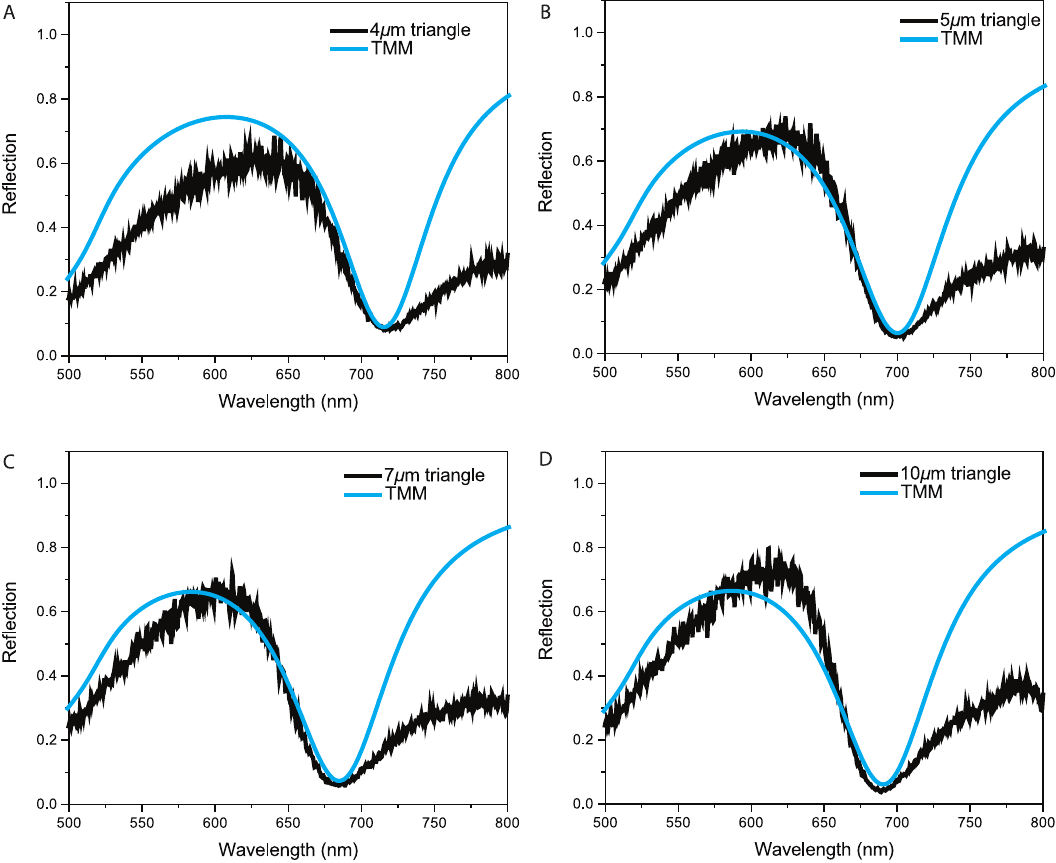}
\caption{\textbf{TMM fitting of the reflecitivty measurements from \Leq=195 nm with different flake sizes.} (\textbf{A-D}) Experimental reflection spectra and their transfer-matrix method (TMM) fittings for various $L_{\rm eq}$. This procedure allows to determine \Leq distance for Fabry-Pérot microcavities shown in Fig.~4F-I of the main text.}
\label{Fig. S3}
\end{figure}

\FloatBarrier

\clearpage

\section{Supplementary Notes}

\vspace{0.25cm}
\noindent
\notetitlelabel{note:tmm}{Determination of the \Leq with Transfer Matrix Method }

In order to determine the \Leq distances for each FP cavity, the reflection spectrum is fitted using the transfer-matrix method (TMM). It is an important parameter since the self-alignment is studied by analyzing its \Leq dependence, systematically. In the TMM fitting for the reflection spectra, the thickness of the cavity mirrors is required as an input parameter. The thickness of the Au seed is 20 nm, in accordance with the nanofabrication process. The thickness of each floating flake is determined from the optical contrast between the seed and the floating flake using optical microscopy. We then utilize this optical contrast, together with a calibrated 20 nm thickness of the seeds, to determine the floating flake thickness. This procedure uses microscope images, collected at the same conditions and the same frame rate as the calibrated seed image. The so-determined thicknesses of the floating flakes (typically in the 20 -- 35~nm range) are subsequently used as input parameters in TMM-calculated reflection spectra of self-assembled FP microcavities.    

\section{Supplementary Videos}
\textbf{Supplementary Video 1.} Accelerated dimer formation for \Leq = 160 nm (video for Fig.~2a,b).\\
\textbf{Supplementary Video 2.} Accelerated dimer formation for \Leq = 110 nm (video for Fig.~2a,b).\\
\textbf{Supplementary Video 3.} Casimir Self-alignment example with both lateral and rotational motion analysis (video for Fig.~1g and Fig.~2c,d).\\
\textbf{Supplementary Video 4.} Variation of $\theta$ with time at various \Leq = 195, 180, 138, 114 nm with edge size 4 $\mu$m seed (video for Fig.~4a-d).\\
\textbf{Supplementary Video 5.} Variation of $\theta$ with time for various edge sizes 4, 5, 7, 10 $\mu$m and with \Leq = 195$\pm$5 nm  (video for Fig.~4f-i).\\

\section{Supplementary Theory}
\subsection{Equations of motion}
\subsubsection{Lateral motion:}
\begin{figure*}[h]
\includegraphics[width=\textwidth]{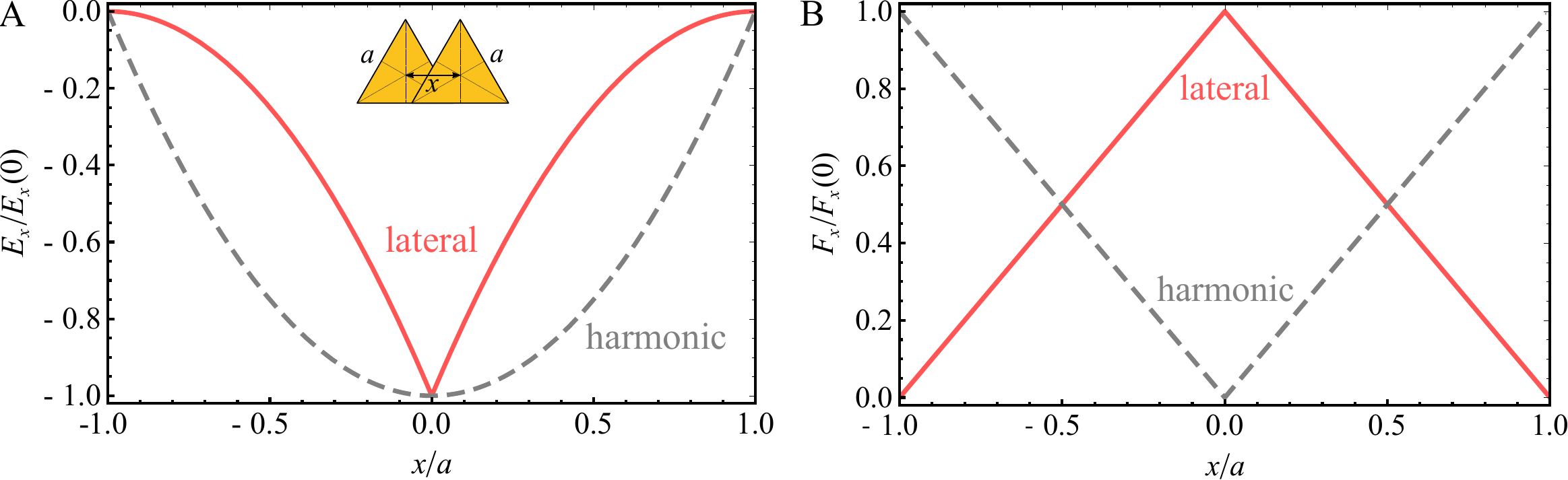}
\caption{\textbf{Qualitative comparison of the lateral motion potential and an arbitrary harmonic one.} (\textbf{A}) Normalized lateral motion potential in the overlap approximation (solid red) and an arbitrary harmonic one (dashed gray) versus the displacement ($x$) between triangles' centers of mass along one of the triangle's edges (see the sketch). (\textbf{B}) The same for the normalized force in the lateral and harmonic cases.}
\label{fig:SI:Theory1}
\end{figure*}

The motion of a floating flake towards the seed of identical size and shape under the influence of a lateral force can be described by the equation of motion for the lateral displacement ($x$) between triangles' centers of mass along one of the triangle's edges:
\begin{equation}
m\ddot{x}=-\gamma_x\dot{x}+F_x(x), \label{motionX}
\end{equation}
where $m$ is the mass of the floating flake, $\gamma_x=k_{\rm B}T/D_x$ is the lateral friction coefficient expressed through the temperature $T$ and lateral diffusion coefficient $D_x$, and $F_x=-\partial E(x)/\partial x$ is the lateral force caused by the lateral potential $E(x)$. In the overlap approximation the lateral potential can be written as $E(x)\approx S(x)U_0(x,L_{\rm eq})$, where $S(x)$ is the overlap area between the flake and the seed, $U_0=U_{\rm C}+U_{\rm e}$ is the total ground-state potential per unit area with the Casimir $U_{\rm C}$ and the electrostatic $U_{\rm e}$ contributions. Importantly, $U_0(L_{\rm eq})<0$, as in any trapping potential at an equilibrium position. For equilateral triangles with the edge size $a$, the overlap area (assuming that the triangles are not rotated with respect to each other) is $S(x)=(a-x)^2\sqrt{3}/4$, which is equal to the triangle's area at $x=0$ (maximum overlap) and becomes zero at $x=a$ (no overlap). Neglecting the edge effects, one can assume the total potential $U_0$ to be independent of $x$ and write the lateral potential as $E(x)\approx U_0(L_{\rm eq})(a-x)^2\sqrt{3}/4$. Interestingly, this potential has the opposite curvature to the harmonic one, as well as a kink point at $x=0$ (see Fig.~\ref{fig:SI:Theory1}a). Note, that such a form of the lateral potential is typical for other flake geometries, such as disks~\cite{Nature2021}. 

\begin{figure}[h]
\begin{center}
\includegraphics[width=0.6\textwidth]
{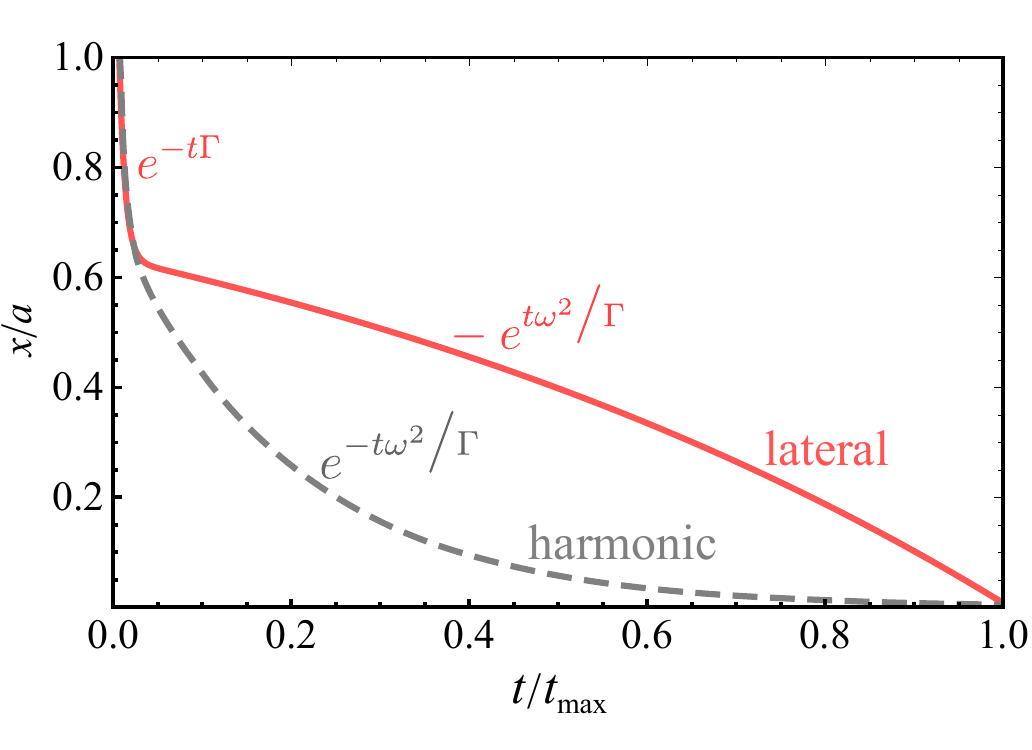}
\end{center}
\caption{\textbf{Qualitative comparison of the solution to the equation of motion for the lateral and arbitrary harmonic potentials.} The solution of the equation of the lateral motion in the deep overdamped regime obtained in the overlap approximation consists of a fast-decaying exponent and a slow one (the parameters are exaggerated for better visibility). While in the harmonic case, the curvature of the slow decaying exponent is positive, in the lateral one it is negative, which gives the decay ending with a non-zero speed (see the discussion in the text).}
\label{fig:SI:Theory2}
\end{figure}

The corresponding lateral force $F_x=(a-x)\sqrt{3}/2\, U_0(L_{\rm eq})$ begins to grow from $x=a$ reaching a maximum at $x=0$, whereas in the harmonic case it is exactly the opposite (see Fig.~\ref{fig:SI:Theory1}b). Thus, accounting for the negativity of $U_0(L_{\rm eq})$, the Eq.~(\ref{motionX}) takes an oscillator equation form with an unusual restoring force:
\begin{equation}
\ddot{x}+\Gamma_x\dot{x}+\omega_x^2(a-x)=0, \label{harmonicX}
\end{equation}
where $\Gamma_x=\gamma_x/m$ is the oscillator lateral damping constant and $\omega_x=\sqrt{|{U_0}|\sqrt{3}\big/2m}$ is the oscillator's eigen frequency. This equation has a general solution of the form: 
\begin{equation}
x(t)=a+C_1e^{-t/2\left(\Gamma_x+\sqrt{\Gamma_x^2+4\omega_x^2}\right)}+C_2e^{-t/2\left(\Gamma_x-\sqrt{\Gamma_x^2+4\omega_x^2}\right)}, \label{Xsolution}
\end{equation}
where the constants $C_{1,2}$ are determined by the boundary conditions. Particularly, from \rev{$x(0)=a$ it follows that $C_1=-C_2$}. The solution (\ref{Xsolution}) differs from that for the harmonic potential only by the positive sign in front of $\omega_x^2$, but this causes a significant difference in the behavior. Firstly, this solution always monotonically decreases and has no oscillation regime, which usually occurs for a harmonic potential when $\Gamma_x<2\omega_x$. Secondly, in the deep overdamped regime when $\Gamma_x>>\omega_x$ the solution (\ref{Xsolution}) reduces to: \rev{
\begin{equation}
x(t)=a+A\left(e^{-t\Gamma_x}-e^{t\omega_x^2/\Gamma_x}\right). \label{Xdamped}
\end{equation}}
Therefore, such a solution contains a fast decaying exponent with large $\Gamma_x$ and \rev{a slow decaying exponent with \textit{inverse} curvature and effective frequency $\omega_x^2/\Gamma_x$ (see Fig.~\ref{fig:SI:Theory2}). The condition $x(t_{\rm max})=0$ gives $A=-a\big/\left(e^{-t_{\rm max}\Gamma_x}-e^{t_{\rm max}\omega_x^2/\Gamma_x} \right)$, or, after simplification, due to $\Gamma_x >> \omega_x^2/\Gamma_x$, $A \approx a e^{-t_{\rm max}\omega_x^2/\Gamma_x}$. So, towards the end of the motion, where only the slow decaying exponent remains, the solution has the form: 
\begin{equation}
x(t)=\widetilde{a}\left(1-e^{(t-t_{\rm max})\omega_x^2/\Gamma_x}\right),
\end{equation}
where $\widetilde{a}=a/(1 - e^{-t_{\rm max}\omega_x^2/\Gamma_x}) > a$ is the renormalized edge size of the triangle, which is close to the actual edge size $a$ as long as $t_{\rm max}>>\Gamma_x/\omega_x^2$. This solution satisfies both boundary conditions: $x(0)=a$ and $x(t_{\rm max}) = 0$.
In the deep overdamped regime, the fast decay occurs on a microsecond time scale and the corresponding exponent leads to a negligible change in the coordinate $x$ as long as $t_{\rm max}>>\Gamma_x/\omega_x^2$. So, on the scale of seconds, on which the measurements are conducted, we can safely neglect the fast exponent as doing so leads to a negligible error in the initial coordinate. The harmonic case, instead, is characterized by a sum of two ordinary decaying exponents with positive curvatures $x(t)=C_1e^{-t\Gamma_x}+C_2e^{-t\omega_x^2/\Gamma_x}$, where $x(0)=a$ results in the relation $C_1+C_2=a$ and $x(\infty)=0$ is always true.}

\begin{figure*}[h]
\includegraphics[width=\textwidth]{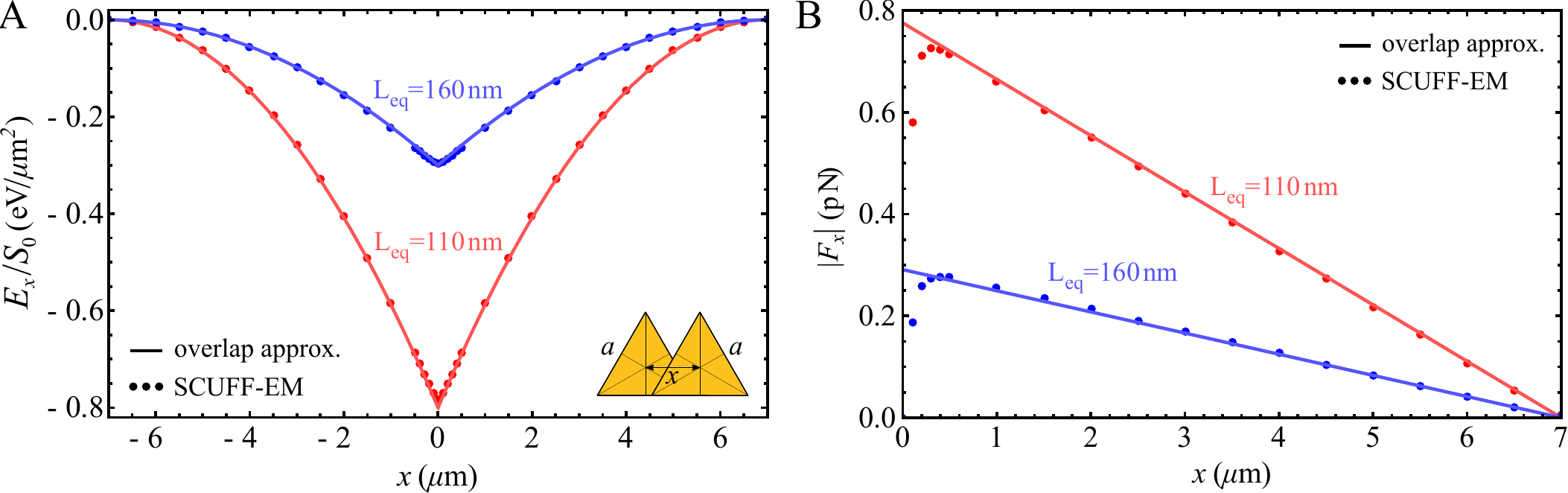}
\caption{\textbf{Numerical verification of the  overlap approximation for the lateral motion.} (\textbf{A}) The lateral motion potential obtained with the overlap approximation (solid lines) and using the SCUFF-EM simulations (points) for the 7-$\mu$m-edge triangles at different $L_{\rm eq}$. (\textbf{B}) The same for the lateral force.}
\label{fig:SI:Theory3}
\end{figure*}

As clearly seen from Fig.~\ref{fig:SI:Theory1}a and Fig.~\ref{fig:SI:Theory2}, the inverse curvature of the lateral potential gives the inverted exponent in the solution. This, in particular, leads to the fact that the lateral motion ends with a non-zero speed, in contrast to the harmonic case. This means that \rev{the flake can pass by the equilibrium position for a short distance, however, because the flake is deeply in the overdamped regime, it will quickly return to the equilibrium position without oscillations}. So, the speed eventually reaches zero due to the damping. If there were no damping, the system would oscillate from $x=a$ to $x=-a$ with the frequency $\omega_x$, even though the solution (\ref{Xsolution}) formally does not contain oscillating functions. However, that is true only in the overlap approximation, which conforms to the special form of the potential having a non-zero derivative at $x=0$ (see Fig.~\ref{fig:SI:Theory1}a). Although this approximation performs surprisingly well, for very small displacements from the equilibrium position, the edge effects result in a harmonic form of the potential (see in Fig.~\ref{fig:SI:Theory3}a; in the vicinity of $x=0$ the points from the SCUFF-EM simulations of the 7-$\mu$m-edge triangles). The role of the edge effects is even more noticeable from the deviation of the force from the linear dependence (see Fig.~\ref{fig:SI:Theory3}b). This harmonicity of the lateral potential near $x=0$, provided by the edge effects, restores the ordinary decaying exponent $e^{-t\omega_x^2/\Gamma_x}$ at the end of the lateral motion \rev{and provides zero final speed}. The harmonicity of the lateral potential near the equilibrium position is even more pronounced for the rotational potential, as we show below.

To estimate the parameters of the solution (\ref{Xsolution}) in our case we can take $D_x \approx 0.05 \mu$m$^2/s$~\cite{schmidt2023tunable} for micro-sized 30-nm-thick Au triangle flakes at room temperature, which results in $\Gamma_x \approx 7$~MHz. Taking the total potential at $L_{\rm eq} = 110$ nm as $U_0 \approx 0.5\,eV/$\um$\!\!^2$~\cite{Nature2021} we obtain $\omega_x \approx 2.4$~kHz, which is approximately three orders of magnitude smaller than $\Gamma_x$. This implies that our system is deep in the overdamped regime, $\Gamma_x >> \omega_x$. Therefore, even if the parameters of the system do slightly deviate from those assumed by us, it still will behave in accordance with the overdamped regime. The system, therefore, behaves according to the solution (\ref{Xdamped}). The fast decaying exponent can be seen at the very beginning of the motion on a sub-microsecond scale, $t < 1/\Gamma_x \sim 0.1$~$\mu$s. The rest of the motion is determined by the slow exponent $-e^{t\omega_x^2/\Gamma_x}$, where $\omega_x^2/\Gamma_x=|{U_0}|\sqrt{3}\big/2\gamma_x=|{U_0}|D_x\sqrt{3}\big/2k_{\rm B}T$. From the estimations above, we arrive at $\omega_x^2/\Gamma_x\sim1$~Hz, which corresponds to the time scale of \emph{seconds} that we indeed observe in our experiments (see e.g. Fig.~2b in the main text and exponential fits of both lateral and rotational motions). Note, that in this regime, the mass goes out of the problem and the motion is determined by the total potential and the diffusion coefficient, which is typical for the Brownian motion. 

\subsubsection{Rotational motion:}
\begin{figure*}[h]
\includegraphics[width=\textwidth]{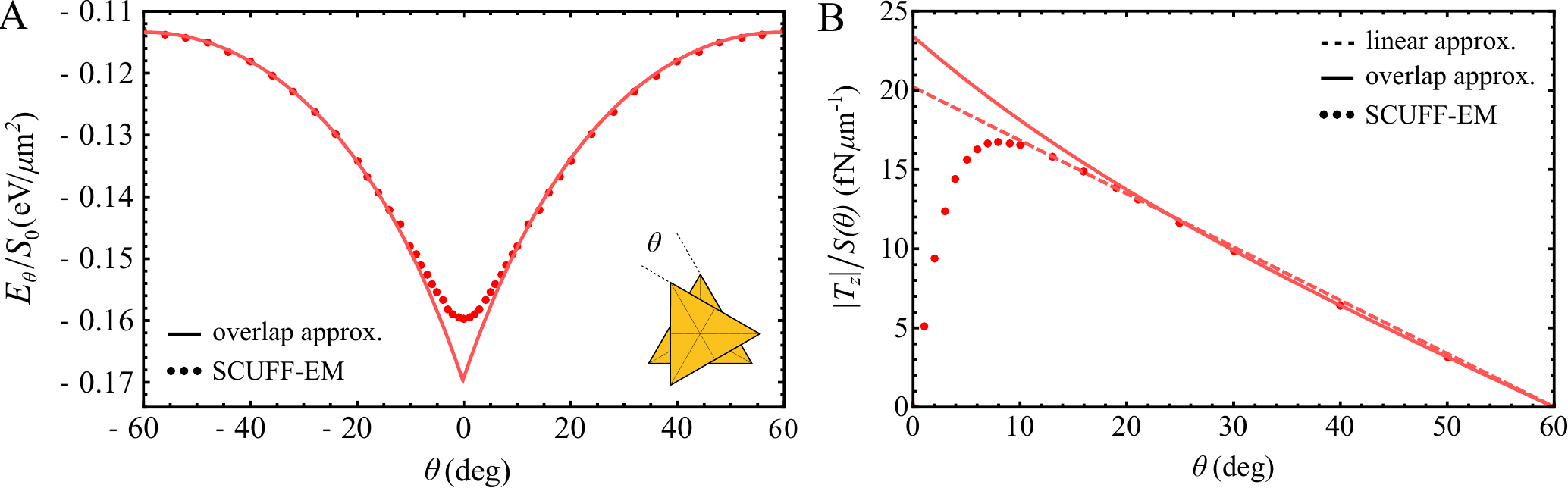}
\caption{\textbf{Numerical verification of the  overlap approximation for the rotation.} (\textbf{A}) The rotational motion potential obtained with the overlap approximation (solid lines) and using the SCUFF-EM simulations (points) for the 6-$\mu$m-edge triangles versus the rotation angle around the center of mass (see the sketch). (\textbf{B}) The same for the torque normalized to the overlap area. Here the dashed line is added, which shows the linear approximation for the normalized torque.}
\label{fig:SI:Theory4}
\end{figure*}

The rotational motion on the angle $\theta$ around an axis normal to a triangle surface and going through its center of mass can be described similarly to the lateral displacements using the equation: 
\begin{equation}
I\ddot{\theta}=-\gamma_\theta\dot{\theta}+T_z(\theta), \label{motionth}
\end{equation}
where $I=ma^2/12$ is the moment of inertia around the center of mass of equilateral triangle with a side $a$, $\gamma_\theta=k_{\rm B}T/D_\theta$, and $T_z=-\partial \big[S(\theta)U_0(\theta,L_{\rm eq})\big]/\partial \theta$ is the torque around the considered axis of rotation defined in the overlap approximation by the overlap area $S(\theta)$ and the total potential per unit area $U_0(\theta,L_{\rm eq})<0$. Note that the units of $D_\theta$ differ from that of $D_x$ by an amount of area. For the rotation of the upper equilateral triangle around its center of mass, the overlap area with the identical lower stationary triangle can be found as follows:
\begin{equation}
S(\theta)=S_0\frac{\rm{tg}(\pi/3-\theta/2)-\rm{tg}(\theta/2)}{\sqrt{3}}, \label{Stheta}
\end{equation}
where $S_0$ is the equilateral triangle’s area. Neglecting the edge effects, one can assume the total potential $U_0$ to be independent of $\theta$ and write the lateral potential as $E(\theta)\approx S(\theta)U_0(L_{\rm eq})$. Plotting this potential with the Eq.~(\ref{Stheta}) clearly shows a perfect correspondence of the overlap approximation to the SCUFF-EM simulations of the 6-$\mu$m-edge triangles at $\theta>10^{\circ}$ (see in Fig.~\ref{fig:SI:Theory4}a). At smaller angles, the edge effects make this potential harmonic, which in this rotational case significantly deviates from the overlap approximation. This leads to even more pronounced deviations in the torque per overlap area, starting already from $\theta\sim20^{\circ}$ (see Fig.~\ref{fig:SI:Theory4}b). In general, Eq.~(\ref{motionth}) with the area given by Eq.~(\ref{Stheta}) is nonlinear, but for the angles $\theta>20^{\circ}$, for which the overlap approximation results in similar torque as the SCUFF-EM simulations, the torque can be approximated by the linear function $T_z(\theta)\approx(\pi/3-\theta)\frac{9}{4\pi}S_0U_0(L_{\rm eq})$ (see dashed line in Fig.~\ref{fig:SI:Theory4}b). 

In this approximation the Eq.~(\ref{motionth}) becomes similar to the Eq.~(\ref{harmonicX}) for the lateral displacements:   
\begin{equation}
\ddot{\theta}+\Gamma_\theta\dot{\theta}+\omega_\theta^2(\pi/3-\theta)=0, \label{harmonicTh}
\end{equation}
where $\Gamma_\theta=\gamma_\theta/I$ and $\omega_\theta=\sqrt{|{U_0}|\frac{9}{4\pi}S_0/I}$. To estimate these parameters we can take $D_\theta\sim D_x/S_0$, thus $\Gamma_\theta=k_{\rm B}T\big/D_\theta I\sim k_{\rm B}T/D_x\cdot S_0/I$. Note that $S_0/I=\frac{a^2\sqrt{3}}{4}\big/\frac{ma^2}{12}=\frac{3\sqrt{3}}{m}$. Therefore, $\Gamma_\theta\sim3\sqrt{3}\Gamma_x\approx5\Gamma_x$ and $\omega_\theta=\sqrt{|{U_0}|\frac{9}{4\pi}\frac{3\sqrt{3}}{m}}=\omega_x\sqrt{27\big/2\pi}\approx2\omega_x$. This implies that for the rotational motion, we obtain the parameters, which, by the order of magnitude, are similar to the lateral displacements. Therefore, the rotation, similarly to translations, occurs on the time scale of \emph{seconds} and adopts the slow exponent form $\theta(t) \sim -e^{\omega_\theta^2/\Gamma_\theta t}$ with $\omega_\theta^2/\Gamma_\theta \sim \frac{4}{5}\omega_x^2/\Gamma_x \sim 1$~Hz, which we indeed observe in the experiment (see Fig.~2d in the main text).

\newpage

\rev{
\subsection{SCUFF-EM calculations}
}

\begin{figure}[h]
\includegraphics[width=1\textwidth]{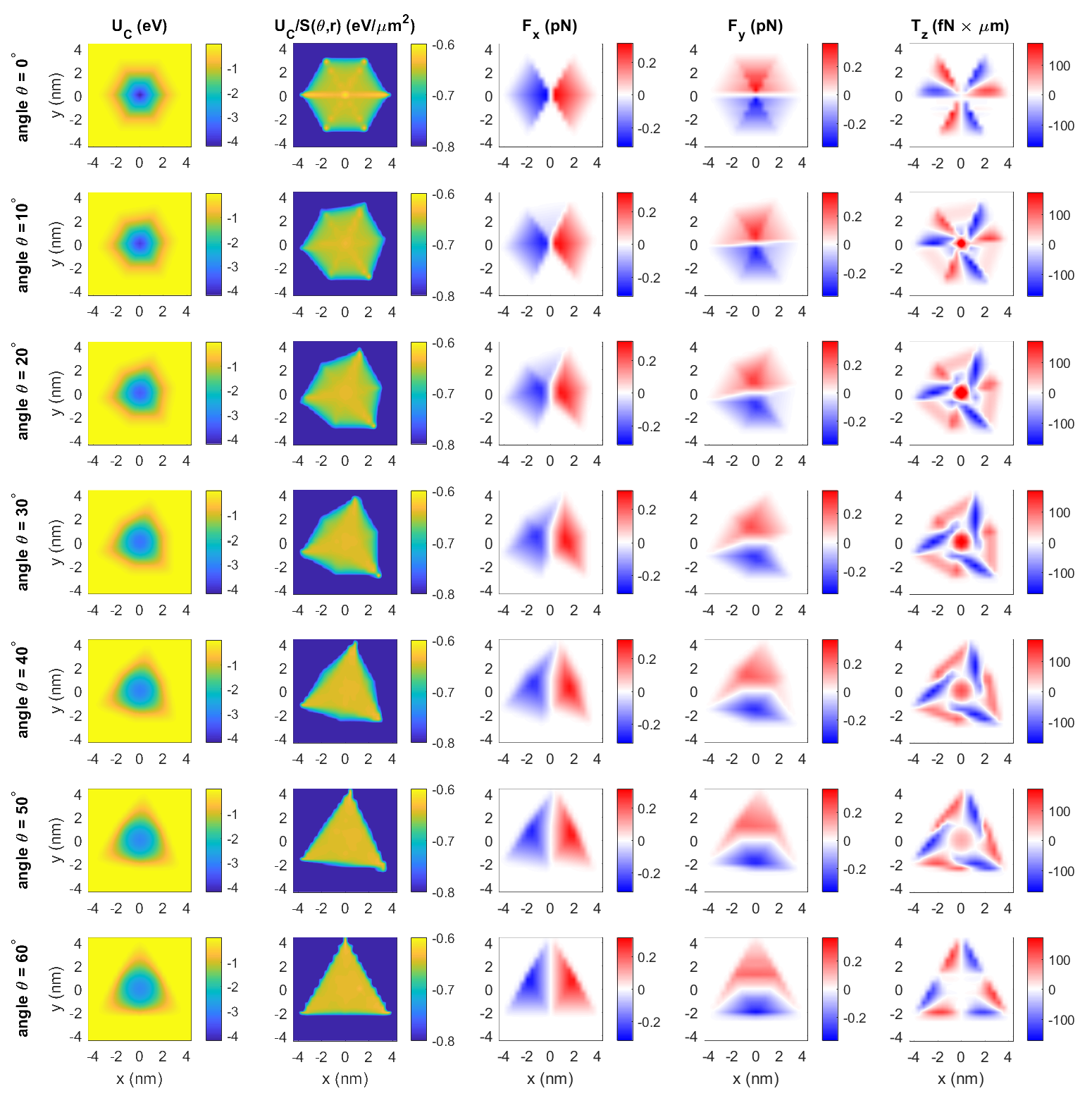}
\caption{\textbf{Calculation of the Casimir potential for finite nanotriangles.} SCUFF-EM simulation results for the relevant parameters of the lateral Casimir interactions of two \unit[40]{nm}-thick gold equilateral triangles, vertically trapped by the balance of the Casimir and electrostatic forces at \Leq= 200 nm and immersed in aqueous solution, depending on the relative shift in $x$ and $y$ in the range -4.4 to 4.4 $\mu$m and rotation angle $\theta$ from 0$^\circ$ to 60$^\circ$; each row corresponds to a particular angle $\theta$. Column 1: total Casimir energy $U_{\rm C}$, Column 2: Casimir energy normalized to surface overlap $U_{\rm C}/S(\theta,r)$, Columns 3 and 4: in-plane Casimir forces $F_x$ and $F_y$, respectively, Column 5: Casmir torque $T_z$ around an axis normal to the triangle surface and going through the center of mass.}
\label{Fig.S_all_angles}
\end{figure}

\FloatBarrier

\clearpage

\rev{
\subsection{Substrate thickness tuning}
}

\begin{figure*}[h]
\begin{center}
\includegraphics[width=0.6\textwidth]{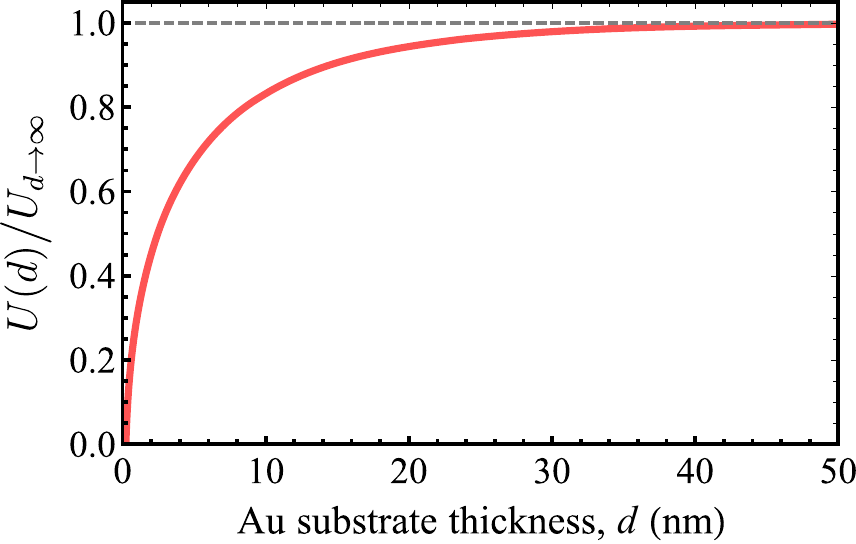}
\end{center}
\caption{\rev{\textbf{Substrate thickness dependence.} The dependence of the Casimir potential on the thickness of the gold substrate. The thickness of the gold floating flake is 30 nm and \Leq = 100 nm.}}
\label{fig:SI:Theory5}
\end{figure*} 

\end{document}